\documentclass[final,3p,times]{elsarticle}

\usepackage{amssymb}

\usepackage{amssymb}
\usepackage[]{amsmath}
\usepackage{graphics}
\usepackage{amssymb}
\usepackage[]{amsmath}
\usepackage{amsthm}
\usepackage{setspace}
\usepackage{epsfig}
\usepackage{subfigure}
\usepackage{extarrows}
\def\[{\begin{equation}}
\def\]{\end{equation}}
\biboptions{numbers,sort&compress}
\usepackage[dvipdfm,colorlinks,linkcolor=red,anchorcolor=blue,citecolor=green]{hyperref}

\usepackage{diagbox}
\usepackage{makecell}
\usepackage{multirow}

\usepackage{amsmath}
\usepackage{times}
\usepackage{anysize}
\marginsize{2cm}{2cm}{0cm}{1.6cm}
\usepackage{mathpazo,color}

\selectfont

\journal{******}

\begin{document}
\begin{frontmatter}



\title{Data-driven localized waves and parameter discovery in the massive Thirring model via extended physics-informed neural networks with interface zones}

\author{Junchao Chen  \fnref{label1}  }
\author{Jin Song \fnref{label2,label3} }
\author{Zijian Zhou \fnref{label2,label3} }
\author{Zhenya Yan \fnref{label2,label3} \corref{cor1}}
\ead{zyyan@mmrc.iss.ac.cn}

\cortext[cor1]{Key Laboratory of Mathematics Mechanization, Academy of Mathematics and Systems Science, Chinese Academy of Sciences, Beijing, 100190, China}

\address[label1]{Department of Mathematics, Lishui University, Lishui, 323000, China}

\address[label2]{Key Laboratory of Mathematics Mechanization, Academy of Mathematics and Systems Science, Chinese Academy of Sciences, Beijing, 100190, China}

\address[label3]{School of Mathematical Sciences, University of Chinese Academy of Sciences, Beijing, 100049, China}

\begin{abstract}
In this paper, we study data-driven localized wave solutions and parameter discovery in the massive Thirring (MT) model via the deep learning in the framework of physics-informed neural networks (PINNs) algorithm.
Abundant data-driven solutions including soliton of bright/dark type, breather and rogue wave are simulated
accurately and analyzed contrastively with relative and absolute errors.
For higher-order localized wave solutions, we employ the extended PINNs (XPINNs) with domain decomposition to capture the complete pictures of dynamic behaviors such as soliton collisions, breather oscillations and rogue-wave superposition.
In particular, we modify the interface line in domain decomposition of XPINNs into a small interface zone and introduce the pseudo initial, residual and gradient conditions as interface conditions linked adjacently with individual neural networks.
Then this modified approach is applied successfully to various solutions ranging from bright-bright soliton, dark-dark soliton, dark-antidark soliton, general breather, Kuznetsov-Ma breather and second-order rogue wave.
Experimental results show that this improved version of XPINNs reduce the complexity of computation with faster convergence rate and keep the quality of learned solutions with smoother stitching performance as well.
For the inverse problems, the unknown coefficient parameters of linear and nonlinear terms in the MT model are identified accurately with and without noise by using the classical PINNs algorithm.

\end{abstract}
\end{frontmatter}

\section{Introduction}

The application of deep learning in the form of neural networks (NNs) to solve partial differential equations (PDEs) has recently been explored in the field of scientific machine learning due to the universal approximation and powerful performance of NNs \cite{hornik1989multilayer,dissanayake1994neural,rackauckas2021universal}.
In particular, by adding a residual term from the given PDE to the loss function with the aid of automatic differentiation,
the physics-informed neural networks (PINNs) approach has been proposed to accurately solve forward problems of predicting the solutions and inverse problems of identifying the model parameters from the measured data \cite{raissi2019physics}.
Unlike the traditional numerical techniques such as the finite element method and finite difference method, the PINNs method is a mesh-free algorithm and  requires the relatively small amounts of data \cite{karniadakis2021physics,lu2021deepxde}.
Based on these advantages, the PINNs has also been applied extensively to different types of PDEs, including integro-differential equations \cite{lu2021deepxde}, fractional PDEs \cite{pang2019fpinns}, and stochastic PDEs \cite{zhang2019quantifying,zhang2020learning}.
However, one of the main disadvantages of PINNs is the high computational cost associated with training NNs, which can lead to the poor performance and low accuracy.
Especially, for modelling problems involving the long-term integration of PDEs, the large amount of data will lead to a rapid increase in the cost of training.
To reduce the computational cost, several improved methods based on PINNs have been introduced to accelerate the convergence of models without loss of performance.
The conservative PINNs (cPINNs) method has been established on discrete subdomains obtained after dividing the computational domain for nonlinear conservation laws, where the subdomains are stitched together on the basis of flux continuity across subdomain interfaces \cite{Jagtap2020Conservative}.
The extended PINNs (XPINNs) approach with the generalized domain decomposition has been proposed to any type of PDEs, where the XPINNs formulation offers highly irregular, convex/non-convex space-time domain decomposition thereby reducing the training cost effectively \cite{Jagtap2020Extended}.
The \emph{hp}-variational PINNs (\emph{hp}-VPINNs) framework  based on the nonlinear approximation of NNs and \emph{hp}-refinement via domain decomposition has has been formulated to solved PDEs using the variational principle \cite{Kharazmi2021variational}.

Based on cPINNs and XPINNs, a parallel PINNs (PPINNs) via domain decomposition has recently been developed to effectively address the multi-scale and multi-physics problems \cite{Shukla2021Parallel,meng2020ppinn}.
The augmented PINNs (APINNs) has been proposed with soft and trainable domain decomposition and flexible parameter \cite{hu2022augmented}.
The PINNs/XPINNs algorithm has been employed to approximate supersonic compressible flows \cite{japtap2022physics} and high-speed aerodynamic flows \cite{mao2020physics}, and to quantify the microstructure of polycrystalline nickel \cite{shukla2021physics}.
The highly ill-posed inverse water wave problems governed by Serre-Green-Naghdi equations have been solved by means of PINNs/XPINNs to choose optimal location of sensors \cite{jagtap2022deep}.
The unified scalable framework for PINNs with temporal domain decomposition has been introduced for time-dependent PDEs\cite{penwarden2023unified}.
The multi-stage training algorithm consisting of individual PINNs has recently been presented via domain decomposition only in the direction of time \cite{mo2021data,pu2023complex}.
Indeed, the theoretical analyses of PINNs/XPINNs including convergence and generalization properties have been proved for linear and nonlinear PDEs through the rigorous error estimates \cite{Mishra2022estimates,Mishra2023estimates,Ryck2023error}, and the conditions under which XPINNs improve generalization have been discussed by proving certain generalization bounds \cite{Hu2022when}.
In addition, the activation function is one of the important hyperparameters in NNs, which usually acts on affine transformation.
The activation function possesses the ability to introduce nonlinearity in NNs in order to capture nonlinear features, and popular activation functions include rectified linear units, maxout units, logistic sigmoid, hyperbolic tangent function and so on \cite{jagtap2022survey}.
In order to improve the performance of NNs and the convergence speed of objective function, global and local adaptive activation functions have
been proposed by introducing scalable parameters into the activation function and adding a slope recovery term into the loss function \cite{jagtap2020locally,jagtap2020adaptive}.
Furthermore, a generalized Kronecker NNs for any adaptive activation functions has been established by Jagtap et al \cite{jagtap2022kronecker}.
The comprehensive survey, the applications-based taxonomy of activation functions and the state-of-the-art adaptive activation functions as well as
the systematic comparison of various fixed and adaptive activation functions have been performed in \cite{jagtap2022survey}.
These adaptive activation functions have also been used extensively in XPINNs (for example, on the inverse supersonic flows \cite{japtap2022physics}).

As a special class of nonlinear PDEs, integrable systems possess remarkable mathematical structures such as infinitely many conservation laws, Lax pairs, abundant symmetry, various exact solutions and so on.
For such type of nonlinear system, the PINNs algorithm could benefit greatly from the underlying integrable properties and achieve a more accurate numerical prediction.
Particularly, a rich variety of solutions with local features such as soliton, breather, rogue wave and periodic wave can be derived exactly by using the classical methods for studying integrable equations, including the inverse scattering transformation, the Darboux transformation, the Hirota bilinear method and the B\"{a}cklund transformation.
These solutions potentially provide us with a large number of training samples for numerical experiments in the framework of PINNs.
For example, the classical and improved PINNs method based on explicit solutions has been used to study data-driven nonlinear wave and parameter discovery in many integrable equations, including KdV equation \cite{raissi2019physics,li2020deep}, mKdV equation \cite{li2020deep}, nonlinear Schr\"{o}dinger (NLS) equation \cite{raissi2019physics,pu2021soliton,li2022mix,wu2022prediction}, derivative NLS equation \cite{peng2022pinn,pu2021solving}, Manakov system \cite{mo2021data,pu2022data,fang2022predicting}, Yajima-Oikawa system \cite{pu2023datadriven}, Camassa-Holm equation \cite{wang2021datadriven} and etc.
It is worth noting that a modified PINNs approach developed by introducing conserved quantities from integrable theory into NNs, has been shown to achieve higher prediction accuracy \cite{lin2022two,fang2022data}.
Moreover, the PINNs scheme has been applied to discover B\"{a}cklund transformation and Miura transform \cite{zhou2022databt}, which are two typical transformations in the field of integrable systems.
The PINNs schemes based on Miura transformations have been established as an implementation method of unsupervised learning for predicting new local wave solutions \cite{lin2023physics}.
In addition, the PINN deep learning has been employed to solve forward and inverse problems of the defocusing NLS equation with a rational potential \cite{wang2021datadrivennls}, the logarithmic NLS with $\mathcal{PT}$-symmetric harmonic potential \cite{zhou2021solving}, the $\mathcal{PT}$-symmetric saturable NLS equation \cite{song2021deep} and other nearly integrable models \cite{zhong2022data1,li2021solving,pu2023complex,wangli2023,zhongming2023}. More recently, the deep learning method with Fourier neural operator was used to study solitons of both integrable nonlinear wave equations~\cite{zhongming-ctp-2023} and integrable fractional nonlinear wave equations~\cite{zhongming-csf-2023}.

The massive Thirring (MT) model takes the following form in laboratory coordinates
\begin{equation}\label{mt01}
\begin{array}{l}
	\mathrm{i} u_x + v + u|v|^2 =0,\\
	\mathrm{i} v_t + u + v|u|^2 =0.
\end{array}
\end{equation}
This two-component nonlinear wave evolution model is known to be completely integrable in the sense of possessing infinitely many conserved quantities \cite{Kuznetsov1977complete} and by means of the inverse scattering transform method \cite{Mikhailov1976integrability,Orfanidis1976soliton}.
It reduces to the massless Thirring model \cite{Thirring1958soluble} if the linear terms $u$ and $v$ are ignored.
In field theory, the MT model (\ref{mt01}) corresponds to an exactly solvable example for one-dimensional nonlinear Dirac system arising as a relativistic extension of the NLS equation \cite{Thirring1958soluble}.
In nonlinear optics, both components in Eq.(\ref{mt01}) represent envelopes of the forward and backward waves respectively \cite{Degasperis2015Bragg}.
Indeed, the MT model (\ref{mt01}) is a particular case of the coupled mode equations with self-phase modulation \cite{Degasperis2015Bragg}.
This coupled mode system has been extensively used to describe pulse propagation in periodic or Bragg nonlinear optical media \cite{Winful1982optical1,Joseph1989optical2,Aceves1989optical3,Eggleton1996optical4,Eggleton1999optical5}, deep water waves for a periodic bottom in ocean \cite{Ruban2008Highly,Ruban2008Water} and superpositions of two hyperfine Zeeman sublevels in atomic Bose-Einstein condensates \cite{Zobay1999Creation}.
The Coleman correspondence between the sine-Gordon equation and the MT model has been found and used to generate solutions in \cite{KaupNewell1977Coleman}.
Various solutions of the MT model like soliton, rogue wave and algebro-geometric solutions as well as other integrable properties have been widely studied by using many different techniques such as the Darboux transformation, B\"acklund transformation, Hirota bilinear method and dressing method \cite{Orfanidis,Kawata79,WadatiMT83,KaupLakoba,Villarroel1991,Lee:1994,Shnider84,Alonso:1984,BarashenkovGetmanov:1987,DateMT2,Bikbaev,MTElbeck,guo2017rogueMT,ye2021super,Degasperis2015darboux,Dajun2022FL}.
With the aid of the bilinear Kadomtsev-Petviashvili (KP) hierarchy reduction approach, one of the authors
has systematacially constructed tau-function solutions of the MT model (\ref{mt01}) including bright, dark soliton and breather as well as rogue wave in compact form of determinants \cite{chen2023tau,chen2023rogue}.

In this paper, we aim to explore data-driven localized waves such as soliton, breather and rogue wave, as well as parameter discovery in the MT model (\ref{mt01}) via the deep learning algorithm.
In forward problems of predicting nonlinear wave, we will employ the traditional PINNs framework to learn the simple solutions including one-soliton of brihgt/dark type and first-order rogue wave.
For the solutions of two-soliton, one-breather and second-order rogue wave with complicated structures, it is necessary to enlarge the computational domain to capture the complete pictures of dynamic behaviors such as soliton collisions,  breather oscillations and rogue wave superpositions.
These requirements force us to utilize the XPINNs approach with domain decomposition for the data-driven experiments correspondingly.
More specifically, the computational domain is divided into a number of subdomains along the time variable $t$ when we treat the MT model as a time evolution problem, hence the XPINNs architecture is also regarded as a multi-stage training algorithm in this situation.
In particular, we slightly modify the interface line in the domain decomposition of XPINNs into a small interface zone shared by two adjacent subdomains.
In the training process, the pseudo initial, residual and gradient points are randomly selected in the small interface zones and are subjected to three types of interface conditions for better stitching effect and faster convergence.
In inverse problems of discovering parameter, the classical PINNs algorithm is applied to identify the coefficients of linear and nonlinear terms in the MT model (\ref{mt01}) in the absence and presence of noise.

The remainder of the paper is organized as follows.
In Section 2, we firstly introduce the classical PINNs method and develop the modified XPINNs approach with small interface zones, in which the domain decomposition with sampling points and the schematic of individual PINN with interface conditions are illustrated in detail.
Then dynamic behaviors of data-driven localized waves in the MT model (\ref{mt01}) ranging widely from one-soliton of bright/dark type, bright-bright soliton, dark-dark soliton, dark-antidark soliton, general breather, Kuznetsov-Ma breather and rogue wave of first- and second-order, are presented via the traditional PINNs and modified XPINNs methods.
In Section 3, based on three types of localized wave solution, we discuss data-driven parameter discovery in the MT model (\ref{mt01}) with and without noise via the classical PINNs algorithm.
Conclusions and discussions are given in the last section.


\section{Data-driven localized waves of the MT model}

In this section, we will use PINNs and XPINNs to solve forward problems of the MT model (\ref{mt01}).
More precisely, we will focus on the the MT model (\ref{mt01}) with initial and Dirichlet boundary conditions as follows:
\begin{equation}\label{mt-ib}
\left\{\begin{array}{l}
\begin{split}
	\mathrm{i} u_x + v + u|v|^2 =0,\\
	\mathrm{i} v_t + u + v|u|^2 =0,
\end{split}\ \ \
x\in[L_0,L_1],\ \ t\in[T_0,T_1],
\\
u(x,T_0)=u_0(x),\ \ v(x,T_0)=v_0(x),\ \ x\in[L_0,L_1],
\\
\begin{split}
	u(L_0,t)=u_{lb}(t), \ \ v(L_0,t)=v_{lb}(t),\ \ \\
	u(L_1,t)=u_{ub}(t), \ \ v(L_1,t)=v_{ub}(t),
\end{split}\ \ \ t\in[T_0,T_1].
 \end{array}\right.
\end{equation}
By taking the complex-valued solutions as $u=p+{\rm i}q$ and $v=r+{\rm i}s$ with the real-valued functions $(p,q)$ and $(r,s)$ being the real and imaginary parts of $u$ and $v$ respectively, the MT model (\ref{mt-ib}) is decomposed into four real equations.
According to the idea of PINNs, we need to construct a complex-valued neural network to approximate the solutions $u$ and $v$.
The complex-valued PINNs for the MT model (\ref{mt-ib}) can be defined as
\begin{equation}\label{mt-pinn1}
\begin{array}{l}
f_u:=	\mathrm{i} \hat{u}_x + \hat{v} + \hat{u}|\hat{v}|^2,\\
f_v:=	\mathrm{i} \hat{v}_t + \hat{u} + \hat{v}|\hat{u}|^2.
\end{array}
\end{equation}
By rewriting $\hat{u}=\hat{p}+{\rm i}\hat{q}$ and $\hat{v}=\hat{r}+{\rm i}\hat{s}$ with the real-valued functions $(\hat{p},\hat{q})$ and $(\hat{r},\hat{s})$, the above models are converted into the following real-valued PINNs:
\begin{equation}\label{mt-pinn2}
\begin{array}{l}
f_p:=\hat{p}_x+\hat{s}+\hat{q}(\hat{r}^2+\hat{s}^2),\ \ f_q:=-\hat{q}_x+\hat{r} +\hat{p}(\hat{r}^2+\hat{s}^2), \\
f_r:=\hat{r}_t+\hat{q}+\hat{s}(\hat{p}^2+\hat{q}^2),\ \ f_s:=-\hat{s}_t+\hat{p}+\hat{r}(\hat{p}^2+\hat{q}^2),
\end{array}
\end{equation}
which possess two inputs $(x,t)$ and four outputs $(\hat{p},\hat{q},\hat{r},\hat{s})$.

\subsection{ Methodology}

\subsubsection{Physics-informed neural networks}

Based on the algorithm of PINNs, we firstly establish a fully connected NN of depth $L$ with
an input layer, $L-1$ hidden-layers and an output layer.
In the $l$th hidden-layer, we assume that it possesses $N_l$ neurons,
then the $l$th layer and the previous layer can be connected by using the affine transformation $\mathcal{A}_l$ and the activation function $\sigma$:
\begin{equation}
\textbf{x}^l=\sigma(\mathcal{A}_l(\textbf{x}^{l-1}))=\sigma(\textbf{W}^l\textbf{x}^{l-1}+\textbf{b}^l),
\end{equation}
where the weight matrix and bias vector are denoted by $\textbf{W}^l\in R^{N_l\times N_{l-1}}$, $\textbf{b}^l\in R^{N_l}$ respectively.
Hence, the whole NN can be written as
\begin{equation}
\mathcal{\hat{U}}(\textbf{x};\Theta) = (\mathcal{A}_{L}\circ\sigma\circ\mathcal{A}_{L-1}\circ \cdots \circ\sigma \circ \mathcal{A}_1)(\textbf{x}),
\end{equation}
where $\mathcal{\hat{U}}(\textbf{x};\Theta)$ represents four outputs $(\hat{p},\hat{q},\hat{r},\hat{s})$ for predicting the solutions $(p,q,r,s)$, and $\textbf{x}$ stands for two inputs $(x,t)$.
The set $\Theta=\{\textbf{W}^l,\textbf{b}^l\}^L_{l=1}\in \bar{\mathcal{P}}$ contains the trainable parameters with $\bar{\mathcal{P}}$ being the parameter space.
Together with the initial and boundary value conditions, the loss function, which consists of different types of mean squared error (MSE), is defined as
\begin{equation}\label{loss1}
\mathcal{L}(\Theta)=Loss=\textbf{W}_R MSE_R + \textbf{W}_{IB} MSE_{IB},
\end{equation}
where $\textbf{W}_R$, $\textbf{W}_{IB}$ are the adjustable weights for the residual and the initial-boundary terms respectively.
The terms $MSE_R$ and $MSE_{IB}$ are given by
\begin{equation}
\begin{array}{l}
MSE_R=\frac{1}{N_R}\sum\limits^{N_R}\limits_{i=1}\left(\left|f_p(x^i_R,t^i_R)\right|^2 + \left|f_q(x^i_R,t^i_R)\right|^2
+\left|f_r(x^i_R,t^i_R)\right|^2
+\left|f_s(x^i_R,t^i_R)\right|^2 \right),
\\
MSE_{IB}=\frac{1}{N_{IB}}\sum\limits^{N_{IB}}\limits_{i=1}\left(\left|\hat{p}(x^i_{IB},t^i_{IB})-p^i\right|^2
+ \left|\hat{q}(x^i_{IB},t^i_{IB})-q^i\right|^2
+ \left|\hat{r}(x^i_{IB},t^i_{IB})-r^i\right|^2
+ \left|\hat{s}(x^i_{IB},t^i_{IB})-s^i\right|^2
\right),
\end{array}
\end{equation}
where $\left \{x^i_R,t^i_R\right \}^{N_R}_{i=1}$ is the set of random residual points, $\left \{x^i_{IB},t^i_{IB}\right \}^{N_{IB}}_{i=1}$ is the set of training points randomly selected from the dataset of initial and boundary value conditions with $(p^i,q^i,r^i,s^i)\equiv$$[p(x^i_{IB},t^i_{IB})$,$q(x^i_{IB},t^i_{IB})$,\\$r(x^i_{IB},t^i_{IB})$,$s(x^i_{IB},t^i_{IB})]$.
The $N_{IB}$ and $N_{R}$ represent the number of points in both sets.
In both terms of loss functions, the term $MSE_R$ is used to measure the deviation from the differential
equations of the MT model on the collocation points with the help of automatic differentiation,
while the term $MSE_{IB}$ enforces the given initial-boundary value conditions as a constraint for ensuring the well-posedness of the MT model.
The PINN algorithm is designed to find the optimized parameters in the set $\Theta$ by minimizing the total loss function.
Consequently, the NN outputs generate the approximate solutions $\hat{u}=\hat{p}+{\rm i}\hat{q}$ and $\hat{v}=\hat{r}+{\rm i}\hat{s}$,
which not only obey the differential equations in the MT model, but also satisfy the initial-boundary value conditions in Eq.(\ref{mt-ib}).

\begin{figure}[!htbp]
\centering
\includegraphics[height=2.2in,width=6.0in]{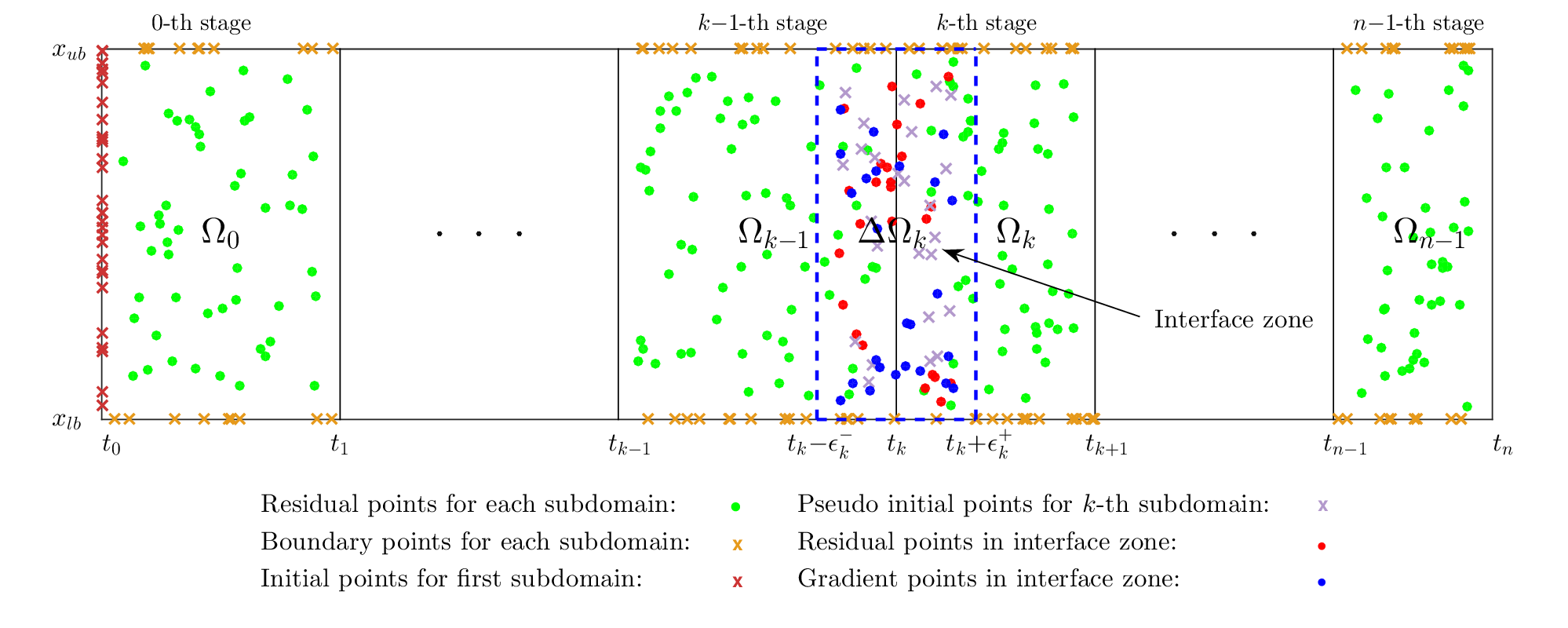}
\caption{The illustrated domain decomposition and distribution of sampling points at each stage.}
\label{frame1}
\end{figure}

\subsubsection{Extended physics-informed neural networks with interface zones}

The basic idea of XPINNs is to divide the computational domain into a number of subdomains and to employ a separate PINN in each subdomain \cite{Jagtap2020Extended}.
Then the training solutions that are subject to the proper interface conditions in the decomposed subdomains are stitched together to obtain a solution over the entire domain.
In XPINNs and cPINNs \cite{Jagtap2020Conservative,Jagtap2020Extended}, a sufficient number of points need to be selected from a common interface line between two adjacent domains to substitute into  the interface conditions.
Here we slightly modify this interface line into a small interface zone that is jointly located in two adjacent domains.
The main reason for this modification is that apart from the pseudo initial and residual conditions, especially the gradient condition will be introduced more reasonably as interface conditions for the smoothness of learned solutions.
This simple improvement will result in better stitching performance and faster convergence as shown below.

To apply XPINNs with small interface zones into our model (\ref{mt-ib}), the whole domain is firstly divided into $N_{s}$ subdomains along the time interval $[T_0,T_1]=[t_0,t_n]=[t_0,t_1]^0\bigcup[t_1,t_2]^1\bigcup\cdots\bigcup[t_{n-1},t_n]^{N_s-1}$.
The $k$-th subdomain $(k{=}0,1,\cdots,N_{s}{-}1{=}n{-}1)$ is described as the set:  $\Omega_k \equiv \{(x,t)|x\in [x_{lb},x_{ub}]=[L_0,L_1], t\in [t_k,t_{k+1}]\}$.
Then the small interface zones are introduced between two adjacent domains.
As our division is along the time variable, the neighborhood of $t_k$ is able to decide the interface zone: $\Delta\Omega_k\equiv\{(x,t)|x\in [x_{lb},x_{ub}], t\in [t_k-\epsilon^-_k,t_k+\epsilon^+_k]\}$ between $k{-}1$-th subdomain and $k$-th one, where $\epsilon^+_{k}$ and $\epsilon^-_k$ are small parameters.
Especially, if $\epsilon^\pm_k=0$, the interface zones reduce to the interface lines of the original XPINNs.
The schematic diagram of such domain decomposition with interface zones is displayed in Fig. \ref{frame1}.

This domain decomposition suggests that $N_{s}$ stages of individual PINNs need to be employed gradually,
where the $k$-th PINN is implemented in the $k$-th subdomain correspondingly.
For the $0$-th stage, the original PINN is used to train the model in the $0$-th subdomain, where the loss function is the same as Eq.(\ref{loss1}) without the additional terms.
The dataset of the initial value is still generated from the initial value condition in the whole model (\ref{mt-ib}), but one
of the boundary value is obtained from the boundary value functions with the time subinterval $t\in [t_0,t_{1}]$.

\begin{figure}[!htbp]
\centering
\includegraphics[height=2.6in,width=4.3in]{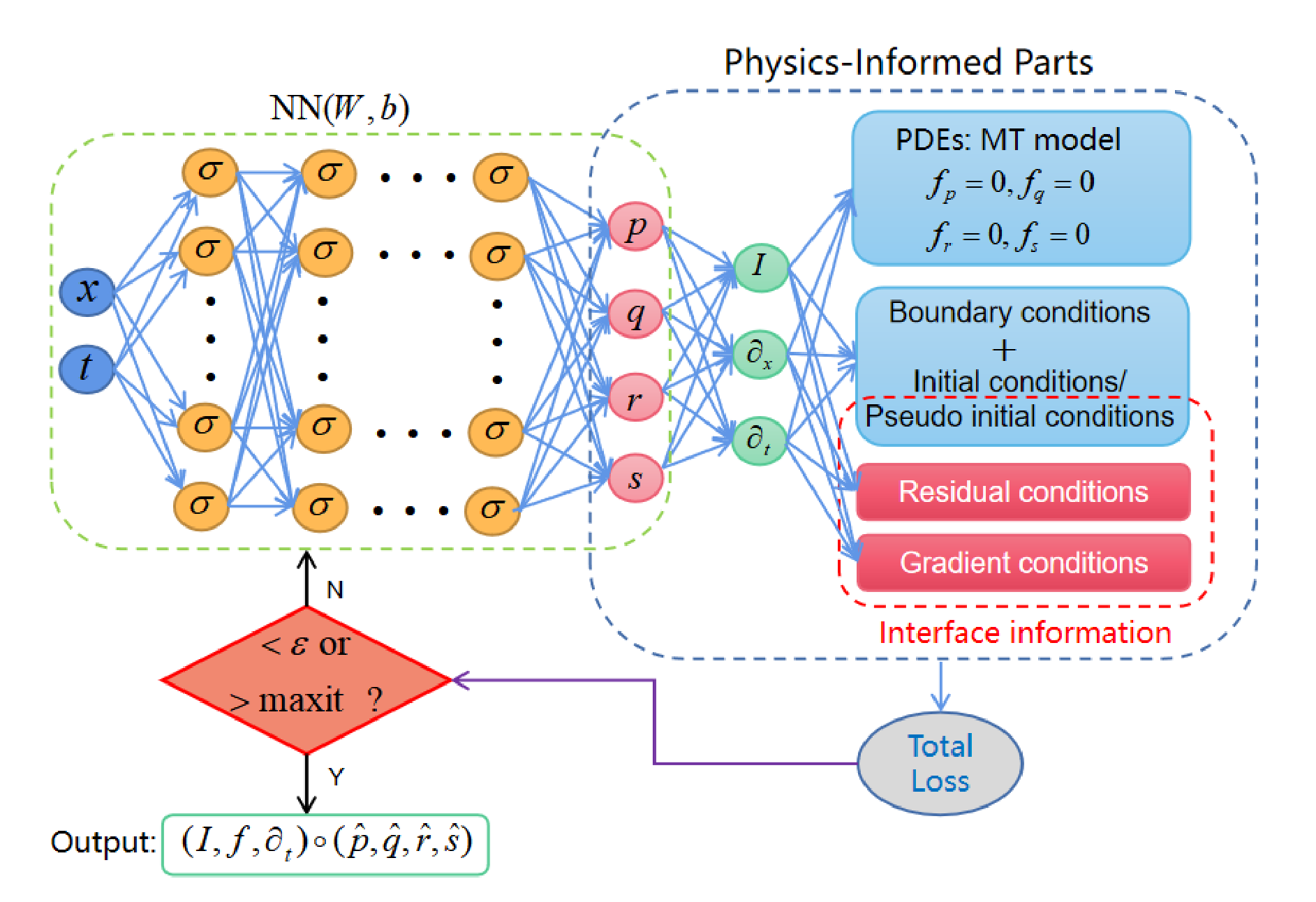}
\caption{The schematic of PINN for the MT model employed at a stage, where NN and
physics-informed network share hyper-parameters with additional loss terms in the interface zone.}
\label{frame2}
\end{figure}

For the $k$-th stage $(k>0)$, the loss function of PINN in the $k$-th subdomain is redefined by adding the interface conditions
\begin{equation}\label{lossk}
\mathcal{L}(\Theta_k)=Loss_k=\textbf{W}^{(k)}_R MSE^{(k)}_R + \textbf{W}^{(k)}_{PIB} MSE^{(k)}_{PIB} +\textbf{W}^{(k)}_{IR} MSE^{(k)}_{IR} + \textbf{W}^{(k)}_{IG} MSE^{(k)}_{IG},
\end{equation}
where $\textbf{W}^{(k)}_R$, $\textbf{W}^{(k)}_{PIB}$, $\textbf{W}^{(k)}_{IR}$  and $\textbf{W}^{(k)}_{IG}$ are the alterable weights for the residual, the pseudo initial-boundary and interface terms where the last two weights correspond to residual and solution smoothness in the interface zone.
The four terms of MSE are expressed as follows:
\begin{equation}
\begin{array}{l}
MSE^{(k)}_R=
\frac{1}{N^{(k)}_R}\sum\limits^{N^{(k)}_R}\limits_{i=1}\left(\left|f_p(x^i_{R_k},t^i_{R_k})\right|^2
+ \left|f_q(x^i_{R_k},t^i_{R_k})\right|^2
+\left|f_r(x^i_{R_k},t^i_{R_k})\right|^2
+\left|f_s(x^i_{R_k},t^i_{R_k})\right|^2 \right),
\\
MSE^{(k)}_{PIB}
= \frac{1}{N^{(k)}_{PIB}}\sum\limits^{N^{(k)}_{PIB}}\limits_{i=1}
\left(\left|\hat{p}(x^i_{PIB_k},t^i_{PIB_k})-p^i_k\right|^2
+  \left|\hat{q}(x^i_{PIB_k},t^i_{PIB_k})-q^i_k\right|^2
+  \left|\hat{r}(x^i_{PIB_k},t^i_{PIB_k})-r^i_k\right|^2
\right.
\\
\left.
\hspace{4cm}
+  \left|\hat{s}(x^i_{PIB_k},t^i_{PIB_k})-s^i_k\right|^2
\right),
\\
MSE^{(k)}_{IR}
= \frac{1}{N^{(k)}_{IR}}\sum\limits^{N^{(k)}_{IR}}\limits_{i=1}
\left(\left|f_p(x^i_{IR_k},t^i_{IR_k})-f^i_{p^-}\right|^2
+  \left|f_q(x^i_{IR_k},t^i_{IR_k})-f^i_{q^-}\right|^2
+  \left|f_r(x^i_{IR_k},t^i_{IR_k})-f^i_{r^-}\right|^2
\right.
\\
\left.
\hspace{4cm}
+ \left|f_s(x^i_{IR_k},t^i_{IR_k})-f^i_{s^-}\right|^2
\right),
\\
MSE^{(k)}_{IG}
= \frac{1}{N^{(k)}_{IG}}\sum\limits^{N^{(k)}_{IG}}\limits_{i=1}
\left(\left|\partial_tp(x^i_{IG_k},t^i_{IG_k})-\partial_t{p^{i-}}\right|^2
+  \left|\partial_tq(x^i_{IG_k},t^i_{IG_k})-\partial_t{q^{i-}}\right|^2
+  \left|\partial_tr(x^i_{IG_k},t^i_{IG_k})-\partial_t{r^{i-}}\right|^2
\right.
\\
\left.
\hspace{4cm}+  \left|\partial_ts(x^i_{IG_k},t^i_{IG_k})-\partial_t{s^{i-}}\right|^2
\right),
\end{array}
\end{equation}
where $\left \{x^i_{R_k},t^i_{R_k}\right \}^{N^{(k)}_R}_{i=1}$ is the set of random residual points in the $k$-th subdomain,
$\left \{x^i_{PIB_k},t^i_{PIB_k}\right \}^{N^{(k)}_{PIB}}_{i=1}$ is the set of training points randomly selected from the dataset of pseudo initial-boundary value conditions.
The values of $(p^i_k,q^i_k,r^i_k,s^i_k)$ contain two parts: If the training points belong to the sub boundary, i.e., $(x^i_{PIB_k},t^i_{PIB_k})\in \{(x,t)|x=x_{lb}, x_{ub}, t\in [t_k,t_{k+1}]\}$, then $p^i_k+{\rm i}q^i_k= u_{lb}(t^i_{PIB_k}),u_{ub}(t^i_{PIB_k})$ and $r^i_k+{\rm i}s^i_k= v_{lb}(t^i_{PIB_k}),v_{ub}(t^i_{PIB_k})$;
If the training points belong to the interface zone, i.e., $(x^i_{PIB_k},t^i_{PIB_k})\in \Delta\Omega_k$, then $(p^i_k,q^i_k,r^i_k,s^i_k)$ are called pseudo initial values obtained from the predicting solution outputs $(\hat{p}^i_{k-1},\hat{q}^i_{k-1},\hat{r}^i_{k-1},\hat{s}^i_{k-1})$ at $(x^i_{PIB_k},t^i_{PIB_k})$ through the $(k-1)$-th PINN with the last optimized parameters.

Furthermore, $\left \{x^i_{IR_k},t^i_{IR_k}\right \}^{N^{(k)}_{IR}}_{i=1}$ and $\left \{x^i_{IG_k},t^i_{IG_k}\right \}^{N^{k}_{IG}}_{i=1}$ denote the set of randomly selected residual points and gradient points respectively in the interface zone $\Delta\Omega_k$.
The $(f^i_{p^-}, f^i_{q^-}, f^i_{r^-}, f^i_{s^-})$ and $(\partial_tp^{i-},\partial_tq^{i-},\partial_tr^{i-},\partial_ts^{i-})$ represent residual and gradient outputs at $(x^i_{IR_k},t^i_{IR_k})$ and $(x^i_{IG_k},t^i_{IG_k})$ respectively from the $k{-}1$-th PINN.
Notice that due to the domain decomposition along the time direction, only the gradients with respect to the variable $t$ need to be computed and then be imposed as one of interface conditions.
The $N^{(k)}_{R}$, $N^{(k)}_{PIB}$, $N^{(k)}_{IR}$ and $N^{(k)}_{IG}$ represent the number of points in the corresponding sets.
The distributions of these sampling points are illustrated in Fig. \ref{frame1}, where the (pseudo) initial-boundary points and residual points for each subdomain are marked with different color crosses and green dots respectively.
In the interface zone $\Delta\Omega_k$, residual and gradient points are distinguished by using red and blue dots.
In fact, we can see clearly that the pseudo initial points are also taken in the interface zone $\Delta\Omega_k$ and
the relevant MSE term can be equivalently seen as one of interface conditions at $k$-th stage.

For four terms in the loss function (\ref{lossk}),  the basic roles of $MSE^{(k)}_R$ and $MSE^{(k)}_{PIB}$ are same as before described in the PINN algorithm on the $k$-th subdomain.
The $MSE^{(k)}_{IR}$ and $MSE^{(k)}_{IG}$ correspond to the residual continuity and $C^1$ solution continuity across the interface zone which is associated with two different sub-networks jointly.
The last two terms and the relevant MSE for pseudo initial points acting as the interface conditions are responsible for transmitting the physical information from the $k{-}1$-th subdomain to the $k$-th one.

To better understand the XPINN algorithm, a schematic of modified PINN in a subdomain is shown in Fig. \ref{frame2}.
Apart from the contributions of the NN part and the physics-informed part, and the loss function is changed by adding certain interface conditions that ensure the quality of stitching and improve the convergence rate.
By minimizing the loss function below a given tolerance $\epsilon$ or up to a prescribed maximum number of iterations,
we can find the optimal values of weights $\textbf{W}$ and biases $\textbf{b}$.
Then the outputs: $(I,f,\partial_t)\circ (\hat{p},\hat{q},\hat{r},\hat{s})$ represent three parts: $I\circ (\hat{p},\hat{q},\hat{r},\hat{s})=(\hat{p},\hat{q},\hat{r},\hat{s})$ are predicting solutions in the subdomain,
$f\circ (\hat{p},\hat{q},\hat{r},\hat{s})=(f_{\hat{p}},f_{\hat{q}},f_{\hat{r}},f_{\hat{s}})$  and $\partial_t \circ (\hat{p},\hat{q},\hat{r},\hat{s})= (\partial_t\hat{p},\partial_t\hat{q},\partial_t\hat{r},\partial_t\hat{s})$ are the residuals and gradients
on random points from the common interface zone.
Finally, these predicting solutions in each subdomain are stitched together and the training errors are measured by the relative $L^2$-norm errors:
\begin{equation*}
{\mbox Error}_{u}=\frac{\sqrt{\sum\limits^{N}\limits_{i=1}|u_i-\hat{u}_i|^2}}{\sqrt{\sum\limits^{N}\limits_{i=1}|u_i|^2}},\qquad
{\mbox Error}_{v}=\frac{\sqrt{\sum\limits^{N}\limits_{i=1}|v_i-\hat{v}_i|^2}}{\sqrt{\sum\limits^{N}\limits_{i=1}|v_i|^2}}.
\end{equation*}

In the above PINN algorithm, the Adam and limited-memory Broyden-Fletcher-Goldfarb-Shanno (L-BFGS) optimizers are employed to train parameters for minimizing the loss function,
where the former is one version of the stochastic gradient descent method and the latter is a full-batch gradient descent optimization algorithm with the aid of a quasi-Newton method.
For smooth and regular solutions, the Adam optimizer is first-order accurate and robust, while the L-BFGS optimizer can seek a better solution due to its second-order accuracy.
Therefore, one would first apply the Adam optimiser to achieve a small value of the loss function, and then switch to the L-BFGS optimiser to pursue a smaller loss function~\cite{raissi2019physics,japtap2022physics}.
In addition, weights are provided with Xavier initialization, biases are initialized to zeros and the activation function is selected as the hyperbolic tangent (tanh) function.
Particularly, it is better to take the optimized weights and biases from the previous stage are as the initialization for the next stage in the XPINN algorithm when both networks have same hidden layers and neurons per layer.
Therefore, in the following experiments for several types different data-driven solutions,
we uniformly construct the NN with 7 hidden layers and 40 neurons per layer in each subdomain.
The recommended weights in the loss function are taken as $\textbf{W}_R=\textbf{W}_{IB}=1$ and $\textbf{W}^{(k)}_R=\textbf{W}^{(k)}_{PIB}=1$, $\textbf{W}^{(k)}_{IR}=0.00001$, $\textbf{W}^{(k)}_{IG}=0.0001$ in these numerical simulations.
At each stage, we first use the $5000$ steps Adam optimization ($10000$ steps for first stage) with the default learning rate $10^{-3}$
and then set the maximum iterations of the L-BFGS optimization to $50000$.
It is mentioned that in the L-BFGS optimization the iteration will stop when
$$
 \frac{L_k-L_{k{+}1}}{{\mbox max}\{|L_k|,|L_{k{+}1}|,1\}}\leq \epsilon_m,
 $$
where $L_k$ denotes loss in the $n$th step L-BFGS optimization and
$\epsilon_m$ stands for the machine epsilon.
Here all the code is based on Python 3.9 and Tensorflow 2.7,  and the default float type is always set  to `float64'.

In what follows, we will focus on the data-driven bright, dark soliton and breather as well as rogue wave solutions of the MT model (\ref{mt01}).
These localized solutions of arbitrary $N$ order have been derived exactly in terms of determinants by using the bilinear KP hierarchy reduction technique \cite{chen2023tau,chen2023rogue}.
For these types of solutions, only the data-driven lower-order cases are considered here, and the higher-order cases can be performed similarly, but with more complicated computations.
In particular, we will apply the classical PINN to  the simple one-soliton and first-order rogue wave solutions, and the XPINN to two-soliton, one-breather and second-order rogue wave solutions.

\subsection{Data-driven bright one- and two-soliton solutions}

The formulae of the bright one-soliton solutions of the MT model (\ref{mt01}) are expressed in the form \cite{chen2023tau}
\begin{equation}\label{brightonesol}
u=\frac{{\rm i}\alpha^*_1 e^{\xi_1}}{\omega_1\left[1-\frac{{\rm i}|\alpha_1|^2\omega_1e^{\xi_1+\xi^*_1}}{(\omega_1+\omega^*_1)^2}\right]}, \qquad
v=\frac{\alpha^*_1 e^{\xi_1}}{1+\frac{{\rm i}|\alpha_1|^2 \omega_1e^{\xi_1+\xi^*_1}}{(\omega_1+\omega^*_1)^2}},
\end{equation}
where $\xi _{1}=\omega_{1}x-\frac{1}{\omega_{1}}t+\xi _{10}$ and $\omega_1$, $\alpha_1$, $\xi _{10}$ are arbitrary complex constants. The symbol $^*$ stands for the complex conjugate hereafter.

\begin{figure}[!htbp]
\centering
\subfigure[]{\includegraphics[height=2.7in,width=1.8in]{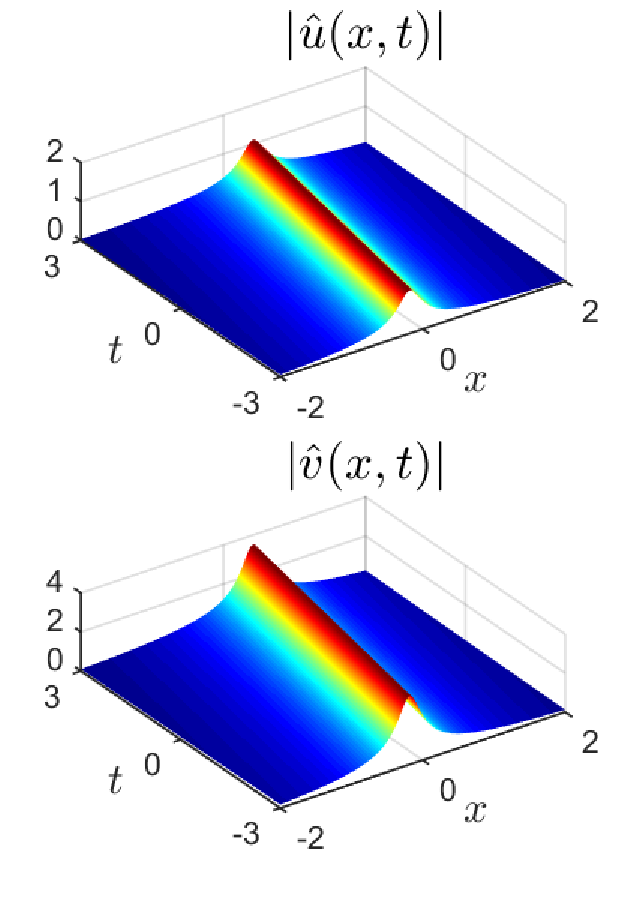}\label{figbrightonea}}\hspace{0.2cm}
\subfigure[]{\includegraphics[height=2.7in,width=1.8in]{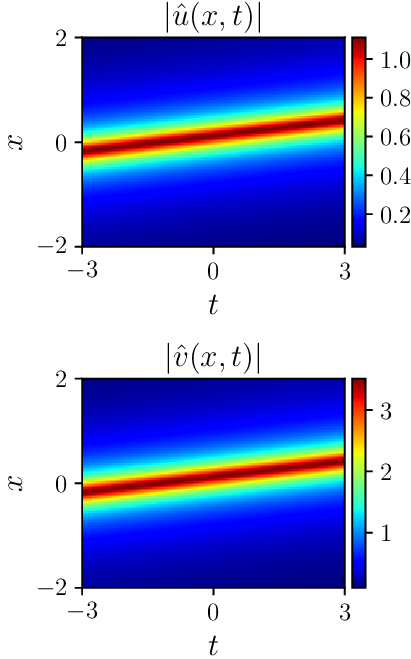}\label{figbrightoneb}}\hspace{0.2cm}
\subfigure[]{\includegraphics[height=2.7in,width=1.8in]{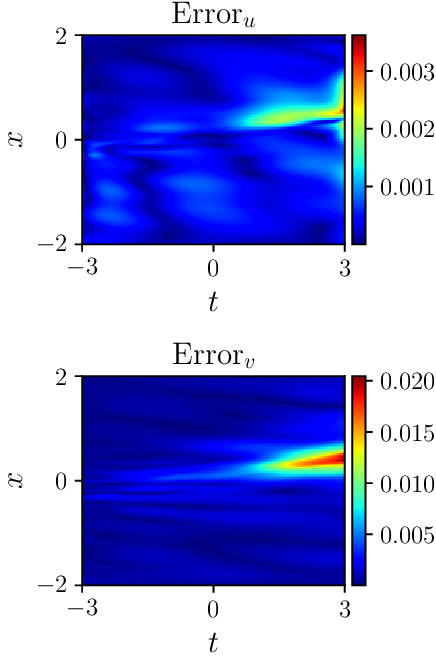}\label{figbrightonec}}
\caption{The data-driven bright one-soliton solution of the MT model (\ref{mt-ib}): (a) and (b) The 3D and 2D density profiles of the learned one-soliton, respectively; (c) The 2D density profiles of the point-wise absolute errors between exact and learned solutions, $\mbox{Error}_u=|u-\hat{u}|$ and $\mbox{Error}_v=|v-\hat{v}|$.
\label{fig-brighone}}
\end{figure}

To simulate the bright one-soliton, the parameters in exact solutions (\ref{brightonesol}) are chosen as $\alpha_1=1$, $\omega_1=1+3{\rm i}$ and $\xi _{10}=0$. The intervals of the computational domain $[L_0,L_1]$ and $[T_0,T_1]$ are taken as $[-2,2]$ and $[-3,3]$, respectively.
Then the initial and Dirichlet boundary conditions in (\ref{mt-ib}) are presented explicitly.
By taking the grid points for the space-time region as $400\times600$ with the equal step length, we get the data sets of the discretized initial-boundary value.
For this simple one-soltion solutions, we employ the original PINN algorithm to conduct the numerical experiment.
Here, $N_{IB}=1000$ training points are randomly selected from the initial-boundary data sets and $N_R=20000$ collocation points are generated in the whole domain by using the Latin Hypercube Sampling (LHS) method.
By optimizing the learnable parameters continually in the NN, we arrive at the learned bright one-solitons $\hat{u}$ and $\hat{v}$.
The relative $L^2$-norm errors for $(|\hat{u}|,|\hat{v}|)$ are $(1.183{\rm e-}03,7.085{\rm e-}04)$.
Figs. \ref{figbrightonea} and \ref{figbrightoneb} exhibit the three-dimensional (3D) structure and two-dimensional (2D) density profiles of
the learned bright one-soliton for the solutions $(|\hat{u}|,|\hat{v}|)$, respectively.
Fig. \ref{figbrightonec} shows the 2D density profiles of the point-wise absolute errors between exact and learned solutions.
The maximum point-wise absolute errors are $(3.621{\rm e-}03, 2.046{\rm e-}02)$, which appear at near the central peaks of soliton with the high gradients.

The expressions for the bright two-soliton solutions of the MT model (\ref{mt01}) are given by
\begin{equation}\label{brighttwosol}
u=\frac{g}{f^{\ast}}\,, \quad
v=\frac{h}{f}\,,
\end{equation}
where tau functions $f$, $g$ and $h$ are defined as
\begin{equation}\label{brighttwosoltaus}
\begin{array}{l}
f=1+c_{1 1^*}e^{\xi _{1}+{\xi}^*_{1}}+c_{21^*}e^{\xi _{2}+{\xi}^*_{1}} +c_{12^*}e^{\xi _{1}+{\xi}^*_{2}}
+c_{22^*}e^{\xi _{2}+{\xi}^*_{2}} +c_{121^*2^*}e^{\xi _{1}+\xi _{2}+{\xi}^*_{1}+{\xi}^*_{2}}, \vspace{0.1in}
\\
g =  \dfrac{\mathrm{i}\alpha^{\ast}_{1}}{\omega_1}e^{\xi _{1}} + \dfrac{\mathrm{i}\alpha^{\ast}_{2}}{\omega_2}e^{\xi_{2}}-\dfrac{\mathrm{i}\omega^*_1}{\omega_1\omega_2} c_{121^*}e^{\xi_{1}+\xi_{2}+{\xi}^*_{1}}-\dfrac{\mathrm{i}k^*_2}{\omega_1\omega_2} c_{122^*}e^{\xi _{1}+\xi_{2}+{\xi}^*_{2}}, \vspace{0.1in}
\\
h=\alpha^*_{1}e^{\xi _{1}}+\alpha^*_{2}e^{\xi _{2}}+c_{121^*}e^{\xi_{1}+\xi_{2}+{\xi}^*_{1}}+c_{122^*}e^{\xi _{1}+\xi_{2}+{\xi}^*_{2}}\,,
\end{array}
\end{equation}
with
\begin{eqnarray*}
&&
c_{ij^*}=\frac{{\rm i}{\omega}_{i} \alpha^*_i\alpha_j}{({\omega}_{i}+{\omega}_{j}^{\ast
})^{2}},\quad
c_{12i^*}=\left( \omega_1-\omega_{2}\right) \omega^*_i \left[ \frac{\alpha^*_{2}c_{1i^*}}{\omega_1(\omega_{2}+{\omega}_{i}^{\ast })}-\frac{\alpha^*_{1}c_{2i^*}}{\omega_2(\omega_{1}+{\omega}_{i}^{\ast })}\right] \,,
\\
&&
c_{121^*2^*}=|\omega_1-\omega_2|^2\left[ \frac{c_{11^*}c_{22^*}}{%
(\omega_{1}+{\omega}_{2}^{\ast })\left( \omega_{2}+{\omega}_{1}^{\ast }\right) }-\frac{c_{12^*}c_{21^*}}{(\omega_{1}+{\omega}_{1}^{\ast })\left( \omega_{2}+{\omega}_{2}^{\ast }\right) }%
\right] \,.
\end{eqnarray*}
Here $\xi _{i}=\omega_{i}x-\frac{1}{\omega_{i}}t+\xi _{i0}$ and $\omega_i$, $\alpha_i$, $\xi _{i0}$ $(i{=}1,2)$ are arbitrary complex constants.

\begin{figure}[!htbp]
\centering
\subfigure[]{\includegraphics[height=1.6in,width=1.8in]{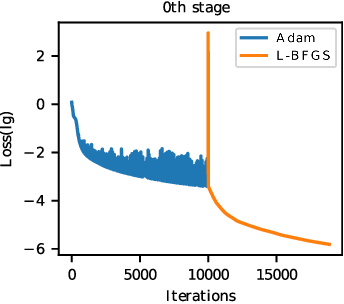}}\hspace{0.2cm}
\subfigure[]{\includegraphics[height=1.6in,width=1.8in]{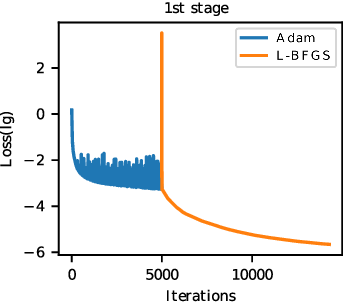}}\hspace{0.2cm}
\subfigure[]{\includegraphics[height=1.6in,width=1.8in]{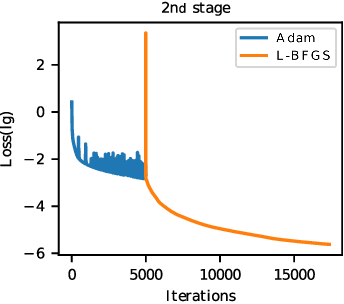}}\hspace{0.2cm}
\subfigure[]{\includegraphics[height=1.6in,width=1.8in]{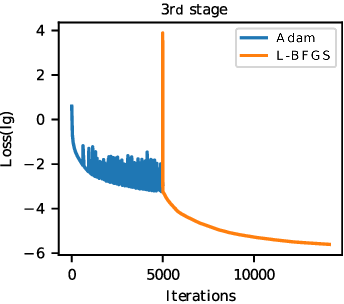}}\hspace{0.2cm}
\subfigure[]{\includegraphics[height=1.6in,width=1.8in]{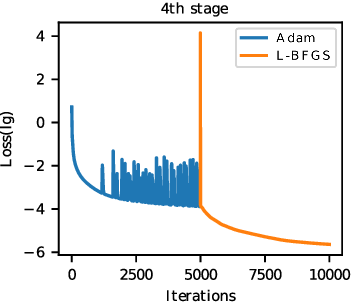}}
\caption{The train loss of the data-driven bright two-soliton solution in each subdomain with Adam (blue) and L-BFGS (orange) optimizations.}
\label{fig-lossplot}
\end{figure}

\begin{figure}[!htbp]
\centering
\subfigure[]{\includegraphics[height=2.7in,width=1.8in]{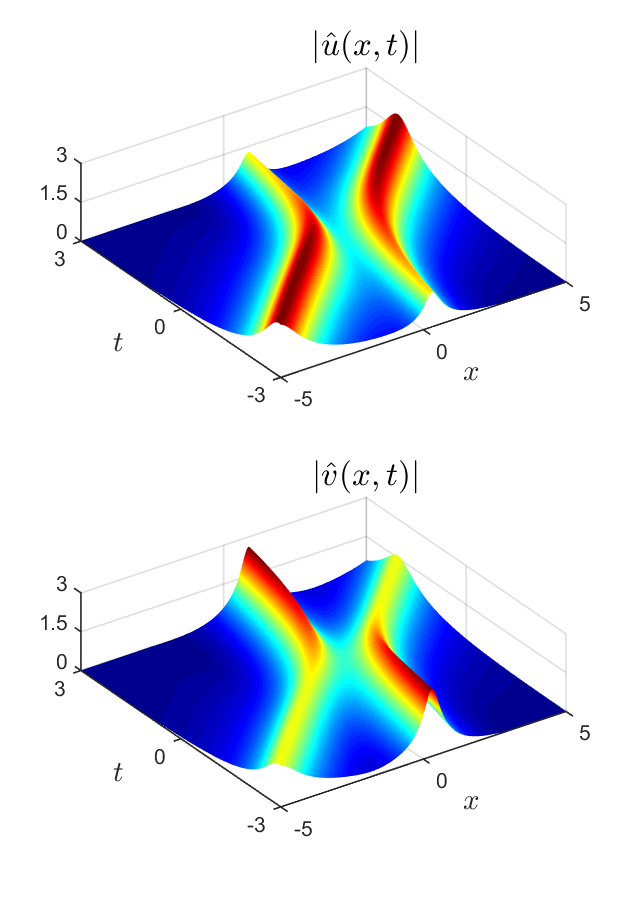}\label{figbrighttwoa}}\hspace{0.2cm}
\subfigure[]{\includegraphics[height=2.7in,width=1.8in]{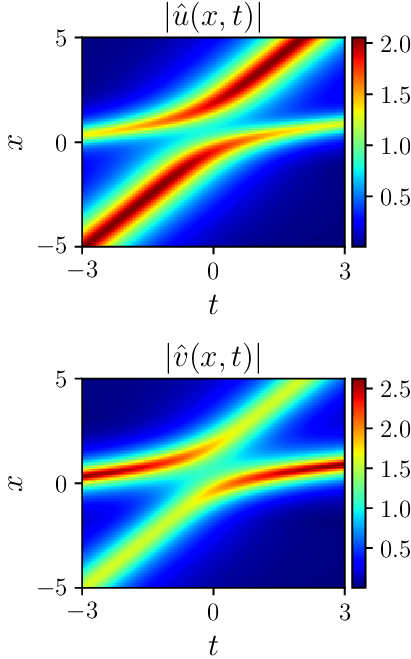}\label{figbrighttwob}}\hspace{0.2cm}
\subfigure[]{\includegraphics[height=2.7in,width=1.8in]{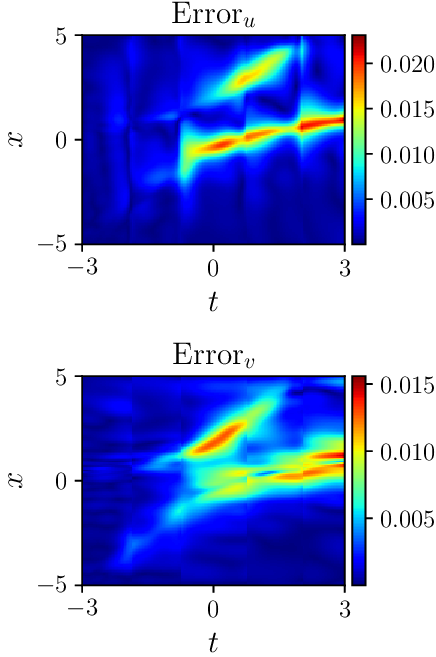}\label{figbrighttwoc}}
\caption{The data-driven bright two-soliton solution of the MT model (\ref{mt-ib}): (a) and (b) The 3D and 2D density profiles of the learned two-soliton, respectively; (c) The 2D density profiles of the point-wise absolute errors between exact and learned solutions, $\mbox{Error}_u=|u-\hat{u}|$ and $\mbox{Error}_v=|v-\hat{v}|$.
\label{fig-brightwo}}
\end{figure}

For the bright two-soliton solutions, the parameters are taken as $\alpha_1=\alpha_2=\frac{3}{2}$, $\omega_1=\frac{3}{2}+\frac{3}{2}{\rm i}$, $\omega_2=\frac{3}{5}+\frac{1}{2}{\rm i}$ and $\xi_{10}=\xi_{20}=0$.
The computational domain $[L_0,L_1]$ and $[T_0,T_1]$ are defined as $[-5,5]$ and $[-3,3]$ respectively, then we have the explicit initial and Dirichlet boundary conditions in (\ref{mt-ib}).
The grid points for the space-time region are set to $600\times1200$ with the equal step length, the data sets of the discretized initial-boundary value can be obtained.
Compared with the bright one-soliton solutions, the bright two-soliton solutions possess the relatively complicated structures.
These features are not well captured by the traditional PINN algorithm, so we must apply the XPINN algorithm to perform this numerical experiment.
The computational domain is divided to 5 subdomains as shown in Table \ref{table-brighttwo}, and $[\epsilon^-_k,\epsilon^+_k]$ are chosen uniformly as $[5,3]\times \frac{3-(-3)}{1200}$ with $k=1,2,3,4$ to decide four small interface zones.
At each stage, the numbers of training points $N^{(k)}_{PIB}$ are given in Table \ref{table-brighttwo} respectively, and the numbers of collocation points $N^{(k)}_{R}$ are set to  $N^{(k)}_{R}=20000$ uniformly.
These points are generated by using the LHS method.
To guarantee the smooth stitching at the line $t=t_k$, it is necessary to take residual points and gradient points with the equal grids in the interface zone $\Delta\Omega_k$.
Hence, we simply take such two types of points as the corresponding part of the grid points in the process of discretising the exact solution.
This implies that here the numbers of both types of sample points are $N^{(k)}_{IR}=N^{(k)}_{IG}=8{\times}600=4800$.
The learned solutions in each subdomain are firstly obtained by training the learnable parameters in each NN.
The loss function plots at each stage are given in Fig. \ref{fig-lossplot}.
Furthermore, by stitching together these predicted solutions in sequence, we have the learned bright two-solitons solutions $\hat{u}$ and $\hat{v}$.
The relative $L^2$-norm errors for $(|\hat{u}|,|\hat{v}|)$ in each subdomain and the whole domain are given in Table \ref{table-brighttwo}.
The 3D structure and 2D density profiles of the learned bright two-soliton for the solutions $(|\hat{u}|,|\hat{v}|)$ in the whole domain are displayed in Figs.~\ref{figbrighttwoa} and \ref{figbrighttwob}, respectively.
The 2D density profiles of the point-wise absolute errors are depicted in Fig. \ref{figbrighttwoc}, in which the maximum values $(2.312{\rm e-}02,1.559{\rm e-}02)$ occur at near the central peaks of soliton.
As we can see from Fig. \ref{figbrighttwoc}, despite the mismatches appear in the high gradient regions, the slight
mismatches along the interface line emerge due to the different sub-networks.
To reduce such mismatch for smooth stitching, one can change the numbers of interface points, the weights of interface terms in the loss function adequately or introduce the higher-order gradient interface conditions.

\begin{table}[htb]
\caption{Relative $L^2$-norm errors of the data-driven bright two-soliton solution in each subdomain and the whole domain with
interface zones.}
\vspace{0.2cm}
	\centering
	\begin{tabular}{cccccccc}
		\hline
		\multirow{2}*{$L^2$ error} & $t$-domain & $[-3,-1.88]$ & $[-1.88,-0.75]$ & $[-0.75,0.75]$ & $[0.75,2]$ & $[2,3]$ & $[-3,3]$  \\
		\cline{2-8}
		~ & $N_{Ib}{+}(N_{PI})$ & 1000 & 2000 & 2500 & 3000 & 3500 \\
		\hline
        $|\hat{u}|$&~&8.887e-04 &9.406e-04 & 2.242e-03 & 2.455e-03 & 3.033e-03  & 1.942e-03\\
        $|\hat{v}|$&~&4.419e-04 &8.044e-04 & 2.307e-03 & 2.613e-03 & 2.742e-03  & 1.992e-03\\
        \hline
	\end{tabular}
\label{table-brighttwo}
\end{table}

\begin{table}[htb]
\caption{Relative $L^2$-norm errors of the data-driven bright two-soliton solution in each subdomain and the whole domain with
interface lines.}
\vspace{0.2cm}
	\centering
	\begin{tabular}{cccccccc}
		\hline
		\multirow{2}*{$L^2$ error} & $t$-domain & $[-3,-1.88]$ & $[-1.88,-0.75]$ & $[-0.75,0.75]$ & $[0.75,2]$ & $[2,3]$ & $[-3,3]$  \\
		\cline{2-8}
		~ & $N_{Ib}{+}(N_{PI})$ & 1000 & 2000 & 2500 & 3000 & 3500 \\
		\hline
        $|\hat{u}|$&~&8.887e-04 &1.072e-03 & 2.192e-03 & 2.576e-03 & 4.179e-03  & 2.101e-03\\
        $|\hat{v}|$&~&4.419e-04 &8.606e-04 & 2.263e-03 & 3.240e-03 & 4.996e-03  & 2.594e-03\\
        \hline
	\end{tabular}
\label{table-brighttwo1}
\end{table}

For the purpose of comparison, we first apply the PINN algorithm directly to learn the bright two-soliton solution.
The network structures and numbers of sample points are taken to be the same as in the numerical experiment for the bright one-soliton.
But the Adam optimization and the maximum iterations of the L-BFGS optimization are adjusted to $20000$ and $50000$ respectively.
The results show that the relative $L^2$-norm errors and the maximum point-wise absolute errors for $(|\hat{u}|,|\hat{v}|)$ reach at $(4.533{\rm e-}01,5.376{\rm e-}01)$ and $(2.951,3.141)$, respectively.
Thus, the original PINN seems to be very poor at predicting the bright two-soliton solution.
On the other hand, we conduct the numerical experiment for the bright two-soliton by using the XPINN algorithm with interface lines, namely $[\epsilon^-_k,\epsilon^+_k]=[0,0]$, where the domain decomposition, all network settings and numbers of sample points are the same as ones of Fig. \ref{fig-brightwo}. For training results,  the relative $L^2$-norm errors for $(|\hat{u}|,|\hat{v}|)$ in each subdomain and the whole domain are listed in Table \ref{table-brighttwo1} and the maximum values of point-wise absolute errors reach at $(2.410{\rm e-}02,2.594{\rm e-}02)$.
Hence, the XPINN algorithm with small interface zones performs slightly better in predicting effects than one with interface lines.

Furthermore, we discuss the influence of the size of small interface zone $\Delta\Omega_k$, which is essentially governed by the thickness $[\epsilon^-_k,\epsilon^+_k]$.
Here we write $[\epsilon^-_k,\epsilon^+_k]=[h^-_k,h^+_k]\times \frac{3-(-3)}{1200}$ with integers $h^-_k$ and $h^+_k$.
In order to preserve the smoothness of stitching at the line $t=t_k$ thoroughly, we need to choose uniform sampling points with the same grid width.
Hence, residual points and gradient points in the interface zone $\Delta\Omega_k$ are taken the same as ones of the grid points in the discretisation of exact solution, which suggests that the numbers of both types of points can be calculated as $N^{(k)}_{IR}=N^{(k)}_{IG}=(h^-_k+h^+_k){\times}600$.
Then, with respect to different size of interface zone $\Delta\Omega_k$, we perform the training experiments for the bright two-soliton, where
the domain decomposition, network architectures, the settings of training step and numbers of sample points are the same as ones in Fig. \ref{fig-brightwo}. According to the results shown in Table \ref{table-differentsizes}, one can see that relative $L^2$ norm errors increase gradually as the thickness of interface zone enlarges.

\begin{table}[htb]
\caption{Relative $L^2$-norm errors of the data-driven bright two-soliton solution in the whole domain with different sizes of interface zones.}
\vspace{0.2cm}
	\centering
	\begin{tabular}{cccccccc}
		 \hline
         $[h^-_k,h^+_k]$ &~& [3,1] & [5,3] & [7,5] & [9,7] & [11,9] & [13, 11] \\
         \hline
		\multirow{2}*{$L^2$ error} & $|\hat{u}|$   & 1.890e-03 & 1.942e-03  &  2.360e-03 &  2.907e-03 &  2.983e-03  & 3.244e-03 \\
		
		~                          & $|\hat{v}|$   & 2.011e-03 & 1.992e-03  &  2.743e-03 &  3.376e-03 &  3.451e-03  & 3.511e-03 \\
		\hline
	\end{tabular}
\label{table-differentsizes}
\end{table}

It is known that the hyperparameters of the networks have a great impact on the performance of NNs, such as width and depth of the network, learning rate, training step and activation function.
As stated above, the activation function is taken as tanh function, and Adam optimization with the $5000$ steps ($10000$ steps for first stage) as well as the L-BFGS optimization with the maximum iterations $50000$ are set uniformly.
Here we mainly consider the influences of width and depth of the network (number of hidden layers and neurons in each layer), and learning rate where the domain decomposition and numbers of sample points remain the same as ones of the experiment for the bright two-soliton in Fig. \ref{fig-brightwo}. These training results are listed in Tables \ref{table-differentlayersneuons} and \ref{table-learningrate}.
It is observed that relative $L^2$ norm errors do not change considerably, and the NN consisting of 7 hidden layers and 40 neurons per layer with   the default learning rate $10^{-3}$ achieves the slightly better network performance in our training experiments.
Particularly, we find that since the L-BFGS optimizer is employed after the Adam optimizer and the L-BFGS optimisation iteration stops automatically at all stages, the final errors do not change much except for the longer training time.

\begin{table}[htb]
\caption{Relative $L^2$-norm errors of the data-driven bright two-soliton solution in the whole domain with different numbers of hidden layers and neurons.}
\vspace{0.2cm}
	\centering
	\begin{tabular}{cccccccc}
		 \hline
         Layers-neurons &~& 5-40 & 5-60 & 7-40 & 7-60  & 9-40 & 9-60 \\
         \hline
		\multirow{2}*{$L^2$ error} & $|\hat{u}|$ &  3.151e-03  & 3.660e-03 & 1.942e-03  &  2.922e-03 &  3.794e-03 &  2.448e-03  \\
		
		~                          & $|\hat{v}|$ &  3.372e-03  & 3.443e-03 & 1.992e-03  &  3.418e-03 &  3.968e-03 &  2.131e-03  \\
		\hline
	\end{tabular}
\label{table-differentlayersneuons}
\end{table}

\begin{table}[htb]
\caption{Relative $L^2$-norm errors of the data-driven bright two-soliton solution in the whole domain with different learning rates in Adam optimization.}
\vspace{0.2cm}
    \centering
	\begin{tabular}{cccccccc}
		 \hline
         Learning rate &~& $8\times10^{-4}$ & $10^{-3}$ & $1.2\times10^{-3}$  \\
         \hline
		\multirow{2}*{$L^2$ error} & $|\hat{u}|$ &  3.322e-03   & 1.942e-03  & 2.766e-03  \\
		
		~                          & $|\hat{v}|$ &  3.265e-03   & 1.992e-03  & 3.132e-03   \\
		\hline
	\end{tabular}
\label{table-learningrate}
\end{table}

\subsection{Data-driven dark one- and two-soliton solutions}

The dark one-soliton solutions of the MT model (\ref{mt01}) are written as \cite{chen2023tau}
\begin{equation} \label{darkonesol}
u=\rho_1 e^{\mathrm{i}\phi} \frac{1-\frac{\mathrm{i} \omega^*_1 K_1} {\omega_1+\omega^*_1}   e^{\xi _{1}+\xi^* _{1}}}{1-\frac{\mathrm{i} \omega^*_1 } {\omega_1+\omega^*_1}e^{\xi _{1}+\xi^*_{1}} }, \qquad
v=\rho_2 e^{\mathrm{i}\phi} \frac{1-\frac{\mathrm{i} \omega^*_1 H_1} {\omega_1+\omega^*_1}   e^{\xi _{1}+\xi^* _{1}}}{1+\frac{\mathrm{i} \omega_1 } {\omega_1+\omega^*_1}e^{\xi _{1}+\xi^* _{1}}},\qquad
\phi=\rho \left(\frac{\rho_2}{\rho_1}x+\frac{\rho_1}{\rho_2}t\right),
\end{equation}
with
\begin{eqnarray*}
&&
K_1=-\frac{\omega_1-\mathrm{i} \alpha}{\omega^*_1+\mathrm{i} \alpha},\qquad
H_1=-\frac{\omega_1- {\rm i}\alpha(1+\rho_1\rho_2)}{\omega^*_1+{\rm i}\alpha(1+\rho_1\rho_2)},
\qquad
\xi_{1}=\frac{\rho_2}{\alpha\rho_1} p_{1}x -\frac{\rho_1  \alpha \rho }{ \rho_2}\frac{t}{\omega_{1}} +\xi_{10}.
\end{eqnarray*}
Here $\alpha$, $\rho_1$, $\rho_2$ are real constants with the restriction that $\rho=1+ \rho_1\rho_2 \ne 0$, $\omega_1$, $\xi_{10}$ are complex constants. These parameters have to satisfy the following algebraic constraint:
\begin{equation*}
|\omega_1-{\rm i}\alpha\rho|^2=\alpha^2\rho_1\rho_2\rho\,.
\end{equation*}

\begin{figure}[!htbp]
\centering
\subfigure[]{\includegraphics[height=2.7in,width=1.8in]{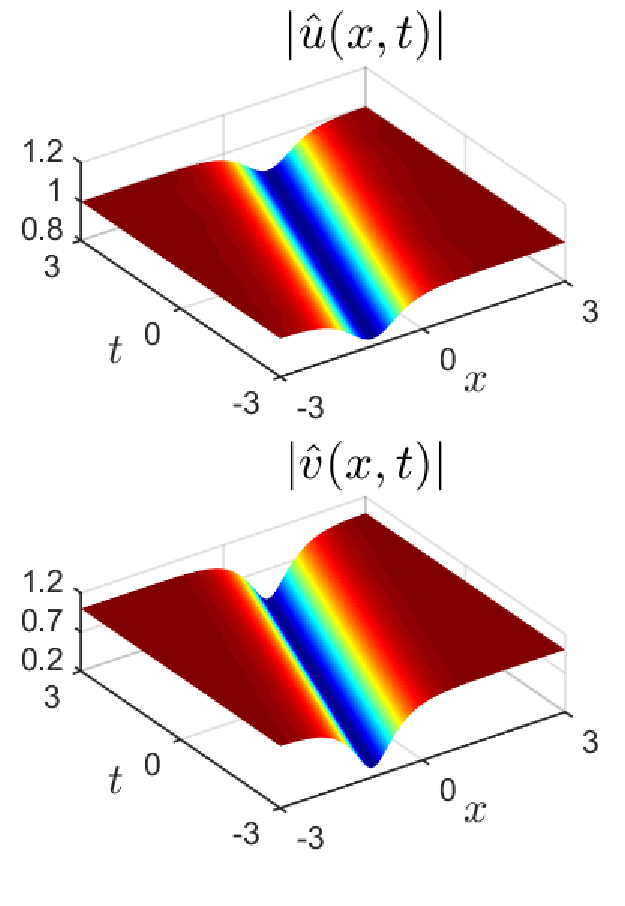}\label{figdarkonea}}\hspace{0.2cm}
\subfigure[]{\includegraphics[height=2.7in,width=1.8in]{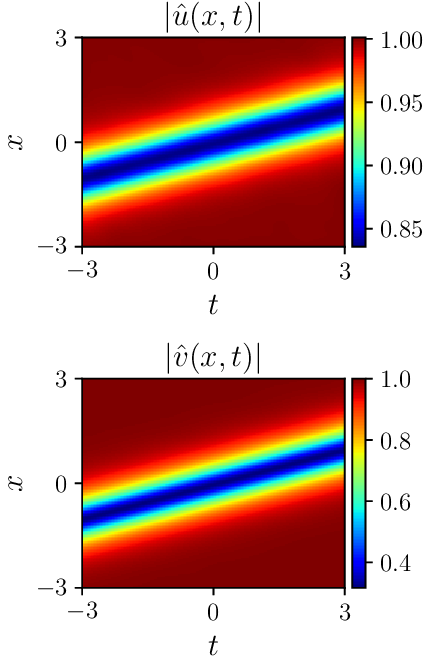}\label{figdarkoneb}}\hspace{0.2cm}
\subfigure[]{\includegraphics[height=2.7in,width=1.8in]{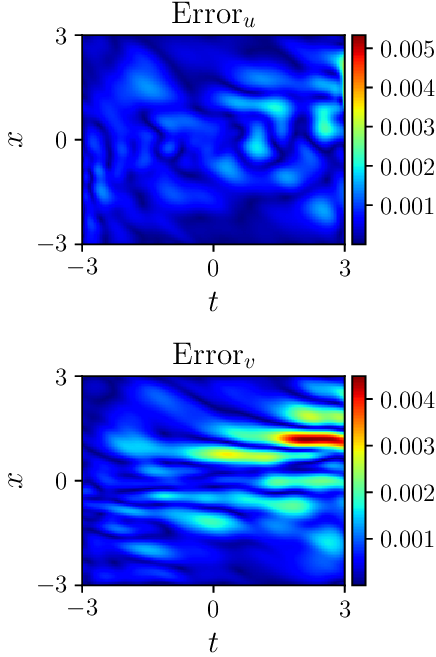}\label{figdarkonec}}
\caption{The data-driven dark one-soliton solution of the MT model (\ref{mt-ib}): (a) and (b) The 3D and 2D density profiles of the learned one-soliton, respectively; (c) The 2D density profiles of the point-wise absolute errors between exact and learned solutions, $\mbox{Error}_u=|u-\hat{u}|$ and $\mbox{Error}_v=|v-\hat{v}|$.
\label{fig-darkone}}
\end{figure}

For the dark one-soliton solutions, the parameter values $\rho_1=\rho_2=\alpha=1$, $\omega_1=-\sqrt{2}+2{\rm i}$ and $\xi_{10}=0$ are chosen for the following numerical simulation.
In the computational domain, the intervals $[L_0,L_1]$ and $[T_0,T_1]$ correspond to $[-3,3]$ and $[-3,3]$.
The initial and Dirichlet boundary conditions in (\ref{mt-ib}) are obtained explicitly from the above exact solutions.
Similarly, the data sets of the discretized initial-boundary value are provided by taking the grid points $400\times600$ with the equal step length in the space-time region.
In this simple case, the classical PINN algorithm is used to conduct the numerical experiment for dark one-soliton solutions.
Furthermore, $N_{IB}=1000$ training points are randomly chosen from the initial-boundary data sets and $N_R=20000$ collocation points are generated in the whole domain by using the LHS method.
After training the learnable parameters in the NN, the predicted dark one-soliton solutions $\hat{u}$ and $\hat{v}$ are successfully acquired.
For this solutions, the relative $L^2$-norm errors for $(|\hat{u}|,|\hat{v}|)$ reach at $(3.998{\rm e-}04,7.319{\rm e-}04)$.
In Figs. \ref{figdarkonea} and \ref{figdarkoneb}, we show the 3D structure and 2D density profiles of
the learned dark one-soliton solutions $(|\hat{u}|,|\hat{v}|)$, respectively.
The 2D density profiles of the point-wise absolute errors are displayed in Fig. \ref{figdarkonec}, in which the maximum point-wise absolute errors $(5.333{\rm e-}03, 4.492{\rm e-}03)$ are found in the high gradient regions near the central peaks of soliton.

The dark two-soliton solutions of the MT model (\ref{mt01}) are expressed as
\begin{equation}\label{darktwosol}
u=\rho_1 e^{\mathrm{i}\phi} \frac{g}{f^{\ast} }, \qquad
v=\rho_2 e^{\mathrm{i}\phi} \frac{{h}}{{f}},\qquad
\phi=\rho \left(\frac{\rho_2}{\rho_1}x+\frac{\rho_1}{\rho_2}t\right),
\end{equation}
where tau functions $f$, $g$ and $h$ are defined as
\begin{equation}\label{darktwosoltaus}
\begin{array}{l}
f= 1+ d_{11^*} e^{\xi_1+\xi^*_1}+ d_{22^*} e^{\xi_2+\xi^*_2}+d_{11^*}d_{22^*}\Omega_{12} e^{\xi_1+\xi_2+\xi^*_1+\xi^*_2},
\vspace{0.1in} \\
g= 1+ d^*_{11^*} K_1 e^{\xi_1+\xi^*_1}+ d^*_{22^*} K_2 e^{\xi_2+\xi^*_2}+d^*_{11^*}d^*_{22^*}K_1K_2 \Omega_{12} e^{\xi_1+\xi_2+\xi^*_1+\xi^*_2}, \vspace{0.1in}
\\
h= 1+ d^*_{11^*} H_1 e^{\xi_1+\xi^*_1}+ d^*_{22^*} H_2 e^{\xi_2+\xi^*_2}+d^*_{11^*}d^*_{22^*}H_1H_2 \Omega_{12} e^{\xi_1+\xi_2+\xi^*_1+\xi^*_2},
\end{array}
\end{equation}
with
\begin{eqnarray*}
&&
d_{ii^*}=\frac{{\rm i}\omega_i}{\omega_i+\omega^*_i},\ \
K_i=-\frac{\omega_i-{\rm i}\alpha}{\omega^*_i+{\rm i}\alpha},\ \
H_i=-\frac{\omega_i-{\rm i}\alpha\rho}{\omega^*_i+{\rm i}\alpha\rho},
\\
&&
\Omega_{12}=\frac{|\omega_1-\omega_2|^2}{|\omega_1+\omega^*_2|^2},\ \
\xi_{i}=\frac{\rho_2}{\alpha\rho_1} \omega_{i}x -\frac{\alpha \rho_1  \rho }{ \rho_2}\frac{t}{\omega_{i}} +\xi_{i0},\ \ i=1,2.
\end{eqnarray*}
Here $\omega_i$, $\xi_{i0}$ ($ i{=}1,2$) are complex constants and $\alpha$, $\rho_1$, $\rho_2$ are real constants with $\rho{=}1+ \rho_1\rho_2 \ne 0$. These parameters need to satisfy the algebraic constraints:
\begin{equation*}
|\omega_i-{\rm i}\alpha\rho|^2=\alpha^2\rho_1\rho_2\rho.
\end{equation*}

\begin{table}[htb]
\caption{Relative $L^2$-norm errors of the data-driven dark-dark soliton solution in each subdomain and the whole domain.}
\vspace{0.2cm}
	\centering
	\begin{tabular}{cccccccc}
		\hline
		\multirow{2}*{$L^2$ error} & $t$-domain & $[-4,-2.5]$ & $[-2.5,-1]$ & $[-1,1]$ & $[1, 2.6]$ & $[2.6,4]$ & $[-4,4]$  \\
		\cline{2-8}
		~ & $N_{Ib}{+}(N_{PI})$ & 1000 & 2000 & 2500 & 3000 & 3500 \\
		\hline
        $|\hat{u}|$&~&1.270e-03&2.582e-03& 2.637e-03 & 4.937e-03 & 7.284e-03  & 4.157e-03\\
        $|\hat{v}|$&~&3.838e-04&8.310e-04& 1.874e-03 & 3.844e-03 & 7.434e-03  & 3.655e-03\\
        \hline
	\end{tabular}
\label{table-darktwo}
\end{table}

\begin{figure}[!htbp]
\centering
\subfigure[]{\includegraphics[height=2.7in,width=1.8in]{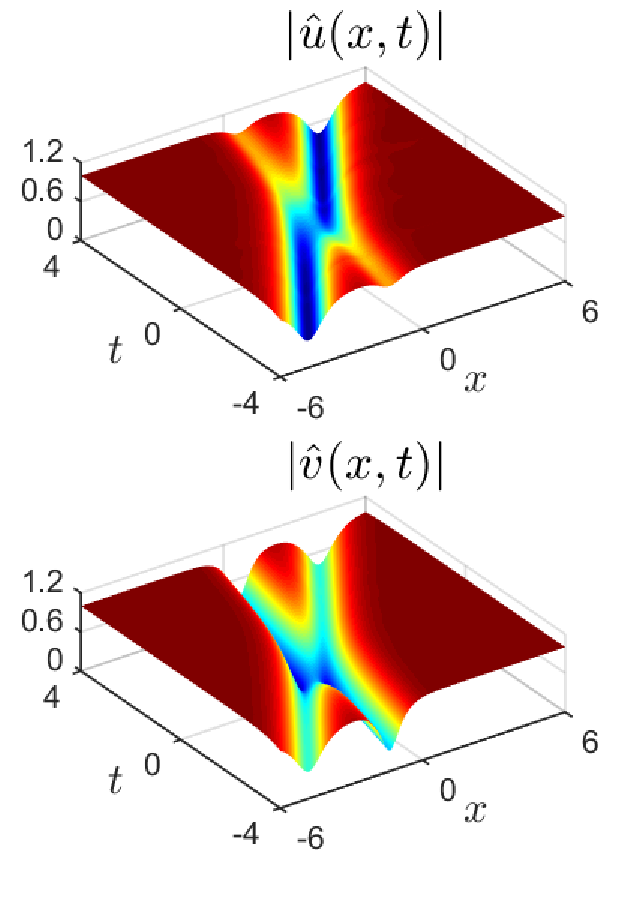}\label{figdarkdarka}}\hspace{0.2cm}
\subfigure[]{\includegraphics[height=2.7in,width=1.8in]{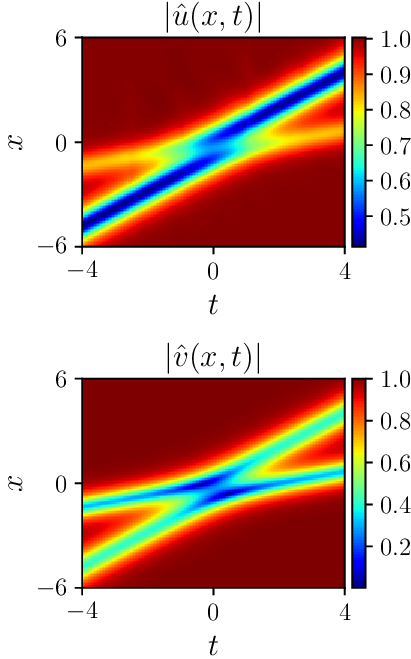}\label{figdarkdarkb}}\hspace{0.2cm}
\subfigure[]{\includegraphics[height=2.7in,width=1.8in]{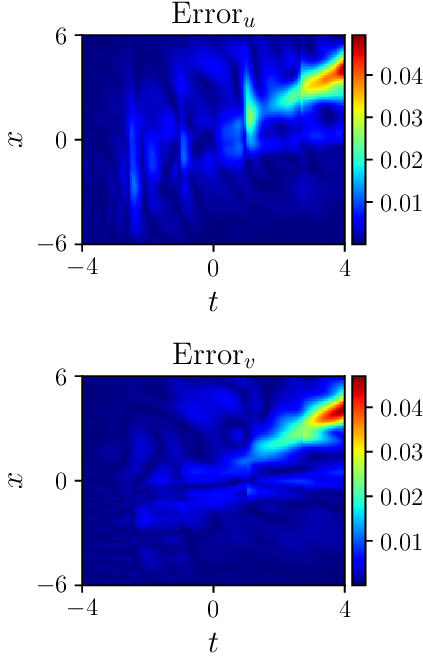}\label{figdarkdarkc}}
\caption{The data-driven dark-dark soliton solution of the MT model (\ref{mt-ib}): (a) and (b) The 3D and 2D density profiles of the learned dark-dark soliton, respectively; (c) The 2D density profiles of the point-wise absolute errors between exact and learned solutions, $\mbox{Error}_u=|u-\hat{u}|$ and $\mbox{Error}_v=|v-\hat{v}|$.
\label{fig-darktwo}}
\end{figure}

Following the detailed analysis of the maximum amplitudes of the dark one-soliton \cite{chen2023tau}, we know that both components $u$ and $v$ allow dark and anti-dark solitons.
Thus the dark two-soliton solutions could support three types of collisions: dark-dark solitons, dark-anti-dark solitons,
and anti-dark-anti-dark ones.
For the purpose of illustration, we will proceed here to learn only two cases: the dark-dark solitons and the dark-anti-dark solitons.
The parameters in both cases are chosen as $\omega_2=-\sqrt{2}+2{\rm i}$, $\xi_{10}=\frac{1}{4}$ for the dark-dark solitons and $\omega_2=\sqrt{2}+2{\rm i}$, $\xi_{10}=-\frac{1}{2}$ for the dark-anti-dark solitons as well as the common values $\rho_1=\rho_2=\alpha=1$, $\omega_1=-1+{\rm i}$, $\xi_{20}=0$.

\begin{figure}[!htbp]
\centering
\subfigure[]{\includegraphics[height=2.7in,width=1.8in]{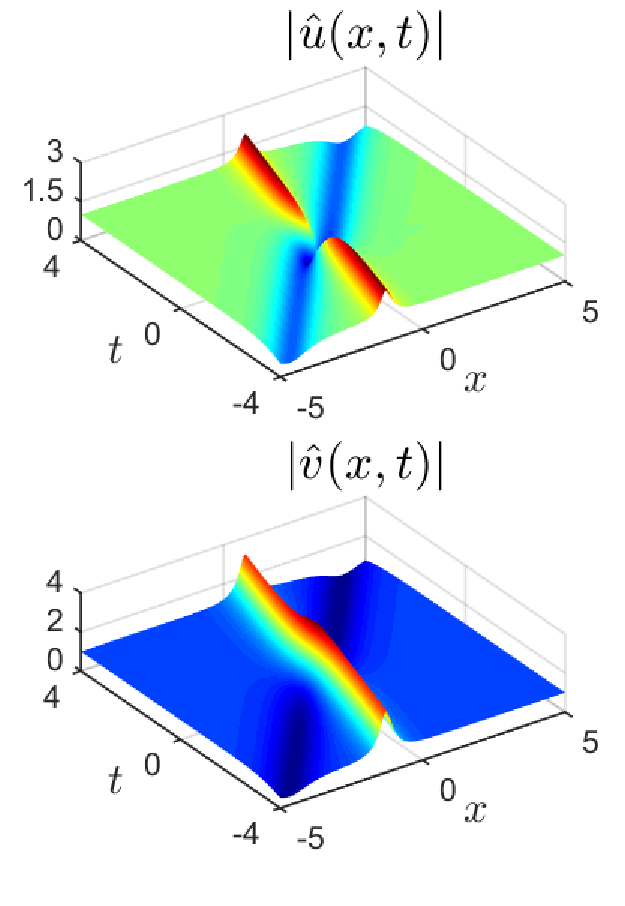}\label{figdarkantidarka}}\hspace{0.2cm}
\subfigure[]{\includegraphics[height=2.7in,width=1.8in]{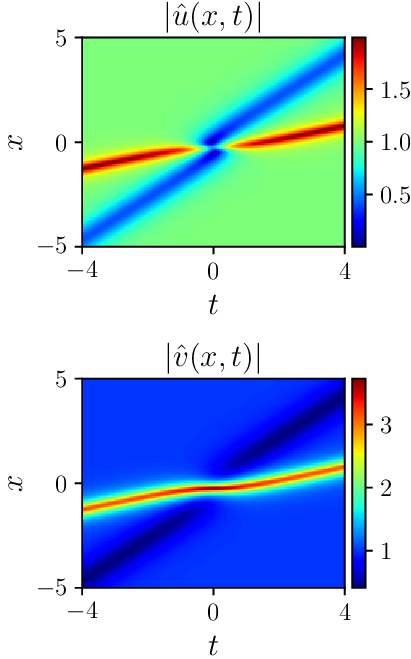}\label{figdarkantidarkb}}\hspace{0.2cm}
\subfigure[]{\includegraphics[height=2.7in,width=1.8in]{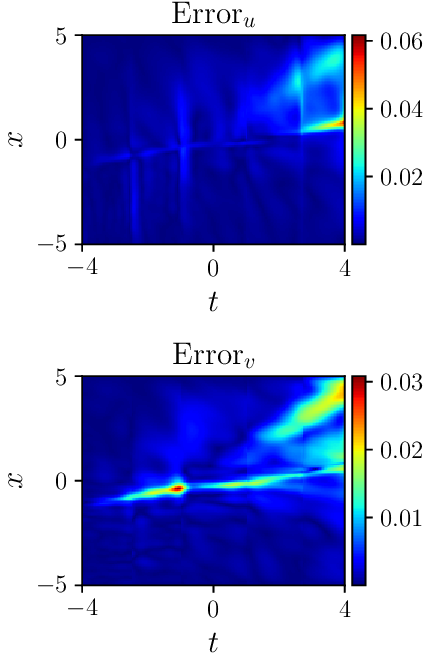}\label{figdarkantidarkc}}
\caption{The data-driven dark-antidark soliton solution of the MT model (\ref{mt-ib}): (a) and (b) The 3D and 2D density profiles of the learned dark-antidark soliton, respectively; (c) The 2D density profiles of the point-wise absolute errors between exact and learned solutions, $\mbox{Error}_u=|u-\hat{u}|$ and $\mbox{Error}_v=|v-\hat{v}|$.
\label{fig-darkantidark}}
\end{figure}

In the following data-driven experiments, the computational domains of both cases $[L_0,L_1]\times[T_0,T_1]$ are taken as $[-6,6]\times[-4,4]$ and $[-5,5]\times[-4,4]$, respectively.
From the definitions given in (\ref{mt-ib}), both types of initial-boundary conditions can be written explicitly.
The data sets of the discretized initial-boundary value are generated by inserting the grid points $600\times1200$ with the equal step length in the space-time regions.
Similar to the bright two-soliton solutions, the dark two-soliton solutions with the relatively complicated structures force us to use the XPINN algorithm for pursuing the better numerical simulation results.
For both cases, we divide the whole computational domain into 5 subdomains as given in Tables \ref{table-darktwo} and \ref{table-darkantidark} respectively, and define $[\epsilon^-_k,\epsilon^+_k]=[5,3]\times \frac{T_1-T_0}{1200}$ uniformly with $k=1,2,3,4$ for the corresponding small interface zones.
In Tables \ref{table-darktwo} and \ref{table-darkantidark}, we list the numbers of training points $N^{(k)}_{PIB}$ at each stage.
The number of collocation points in each subdomain is $N^{(k)}_{R}=20000$ uniformly, which are produced by using the LHS method.
To ensure smooth stitching at the line $t=t_k$, we take residual points and gradient points in each interface zone as the corresponding part of the grid points in the discretisation of exact solution.
Thus the numbers of two types of sample points are $N^{(k)}_{IR}=N^{(k)}_{IG}=4800$.
After training the learnable parameters in each NN and stitching together the predicted solutions in all subdomains,
both types of the dark two-soliton solutions $\hat{u}$ and $\hat{v}$ are obtained in the whole computational domain.
In each subdomain and the whole domain, the relative $L^2$-norm errors for $(|\hat{u}|,|\hat{v}|)$ are calculated in Tables \ref{table-darktwo} and \ref{table-darkantidark}.
For the dark-dark solitons and the dark-anti-dark solitons, the 3D structure and 2D density profiles of the learned solutions as well as the 2D density profiles of the point-wise absolute errors are shown in Figs. \ref{fig-darktwo} and \ref{fig-darkantidark} respectively.
In the comparisons between exact solutions and learned ones for two cases, the maximum absolute errors of the solutions $(|\hat{u}|,|\hat{v}|)$ are $(4.940{\rm e-}02,4.699{\rm e-}02)$ and $(6.169{\rm e-}02,3.082{\rm e-}02)$ near the central peaks and valleys of solitons that correspond to the high gradient regions.
In addition, we can observe that apart from the large errors occurring near the wave peaks and valleys, the adjacent sub-networks with the different  architectures lead to certain errors along the interface line.
This type of mismatch can be reduced by adding more interface information, such as increasing the numbers of interface points and the weights of interface terms in the loss function, and imposing the additional interface conditions.


\begin{table}[htb]
\caption{Relative $L^2$-norm errors of the data-driven dark-antidark soliton solution in each subdomains and the whole domain.}
\vspace{0.2cm}
	\centering
	\begin{tabular}{cccccccc}
		\hline
		\multirow{2}*{$L^2$ error} & $t$-domain & $[-4,-2.5]$ & $[-2.5,-1]$ & $[-1,1]$ & $[1, 2.6]$ & $[2.6,4]$ & $[-4,4]$  \\
		\cline{2-8}
		~ & $N_{Ib}{+}(N_{PI})$ & 1000 & 2000 & 2500 & 3000 & 3500 \\
		\hline
        $|\hat{u}|$&~&6.277e-04&1.468e-03& 1.834e-03 & 2.622e-03 & 4.318e-03  & 2.449e-03\\
        $|\hat{v}|$&~&4.381e-04&8.711e-04& 1.135e-03 & 2.161e-03 & 3.759e-03  & 1.959e-03\\
        \hline
	\end{tabular}
\label{table-darkantidark}
\end{table}

\subsection{Data-driven one-breather solutions}

The one-breather solutions of the MT model (\ref{mt01}) are as follows \cite{chen2023tau}:
\begin{equation}\label{breatheronesol}
u=\rho_1 e^{\mathrm{i}\phi} \frac{g}{f^{\ast} }, \ \
v=\rho_2 e^{\mathrm{i}\phi} \frac{{h}}{{f}}, \ \ \phi=\rho \left(\frac{\rho_2}{\rho_1}x+\frac{\rho_1}{\rho_2}t\right),
\end{equation}
where tau functions $f$, $g$ and $h$ are given by
\begin{equation}\label{breatheronesoltaus}
\begin{array}{l}
f=\frac{e^{\zeta_1+\zeta^*_1}}{\omega_{11}+\omega^*_{11}}\frac{\omega_{11}}{\omega_{21}}
+\frac{e^{\zeta_1}}{\omega_{11}+\omega^*_{21}}\frac{\omega_{11}}{\omega_{21}}
+\frac{e^{\zeta^*_1}}{\omega_{21}+\omega^*_{11}}
+\frac{1}{\omega_{21}+\omega^*_{21}}, \vspace{0.1in}
\\
g=\frac{\Theta_1 \Theta^{*-1}_1 e^{\zeta_i+\zeta^*_j}}{\omega_{11}+\omega^*_{11}}\frac{\omega^*_{11}}{\omega^*_{21}}
+\frac{\Theta_1 e^{\zeta_1}}{\omega_{11}+\omega^*_{21}}
+\frac{\Theta^{*-1}_1 e^{\zeta^*_1}}{\omega_{21}+\omega^*_{11}}\frac{\omega^*_{11}}{\omega^*_{21}}
+\frac{1}{\omega_{21}+\omega^*_{21}}, \vspace{0.1in}
\\
h=\frac{\Lambda_1 \Lambda^{*-1}_1 e^{\zeta_1+\zeta^*_1}}{\omega_{11}+\omega^*_{11}}\frac{\omega^*_{11}}{\omega^*_{21}}
+\frac{\Lambda_1 e^{\zeta_1}}{\omega_{11}+\omega^*_{21}}
+\frac{\Lambda^{*-1}_1 e^{\zeta^*_1}}{\omega_{21}+\omega^*_{11}}\frac{\omega^*_{11}}{\omega^*_{21}}
+\frac{1}{\omega_{21}+\omega^*_{21}},
\end{array}
\end{equation}
with
\begin{eqnarray*}
&&
\Theta_1= \frac{\omega_{11}-{\rm i}\alpha}{\omega_{21}-{\rm i}\alpha} ,\qquad
\Lambda_1=\frac{\omega_{11}-{\rm i}\alpha\rho}{\omega_{21}-{\rm i}\alpha\rho},
\qquad
\zeta_1=(\omega_{11}-\omega_{21})\left(\frac{\rho_2}{\alpha\rho_1}x+\frac{\rho_1\alpha\rho}{\rho_2 \omega_{11}\omega_{21}}t\right)+\zeta_{10}.
\end{eqnarray*}
Here $\omega_{11}$, $\omega_{21}$, $\zeta_{10}$ are complex constants and $\alpha$, $\rho_1$, $\rho_2$ are real constants with $\rho=1+ \rho_1\rho_2 \ne 0$, in which these parameters need to obey the algebraic constraint:
\begin{equation*}
(\omega_{11}-{\rm i}\alpha\rho)(\omega_{21}-{\rm i}\alpha\rho)=-\alpha^2\rho_1\rho_2\rho.
\end{equation*}

\begin{table}[htb]
\caption{Relative $L^2$-norm errors of the data-driven one-breather solution in each subdomain and the whole domain.}
\vspace{0.2cm}
	\centering
	\begin{tabular}{cccccccc}
		\hline
		\multirow{2}*{$L^2$ error} & $t$-domain & $[-7,-4.94]$ & $[-4.94,-2.88]$ & $[-2.88,-0.13]$ & $[-0.13, 2.17]$ & $[2.17,4]$ & $[-7,4]$  \\
		\cline{2-8}
		~ & $N_{Ib}{+}(N_{PI})$ & 1000 & 2000 & 2500 & 3000 & 3500 \\
		\hline
        $|\hat{u}|$&~&4.582e-04 & 2.405e-03 & 4.290e-03 & 3.358e-03 & 3.174e-03  & 3.112e-03\\
        $|\hat{v}|$&~&5.341e-04 & 4.362e-03 & 1.304e-02 & 6.867e-03 & 3.140e-03  & 7.575e-03\\
        \hline
	\end{tabular}
\label{table-breatherone}
\end{table}

\begin{figure}[!htbp]
\centering
\subfigure[]{\includegraphics[height=2.7in,width=1.8in]{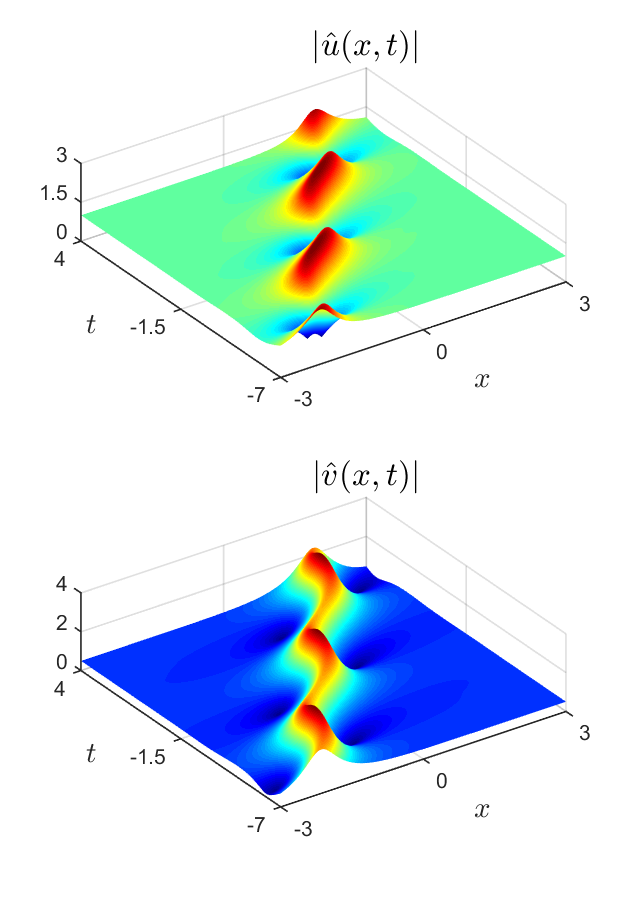}\label{figbreatheronea}}\hspace{0.2cm}
\subfigure[]{\includegraphics[height=2.7in,width=1.8in]{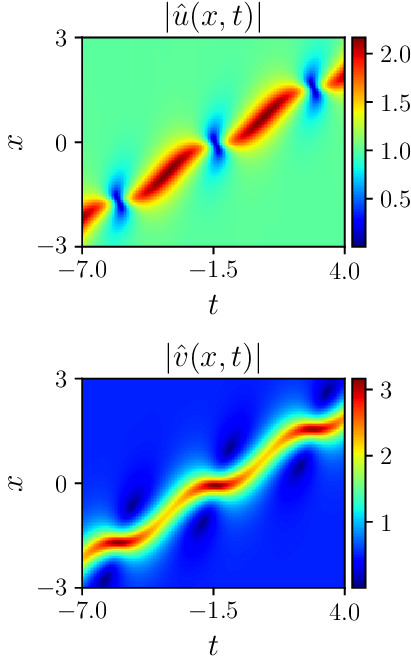}\label{figbreatheroneb}}\hspace{0.2cm}
\subfigure[]{\includegraphics[height=2.7in,width=1.8in]{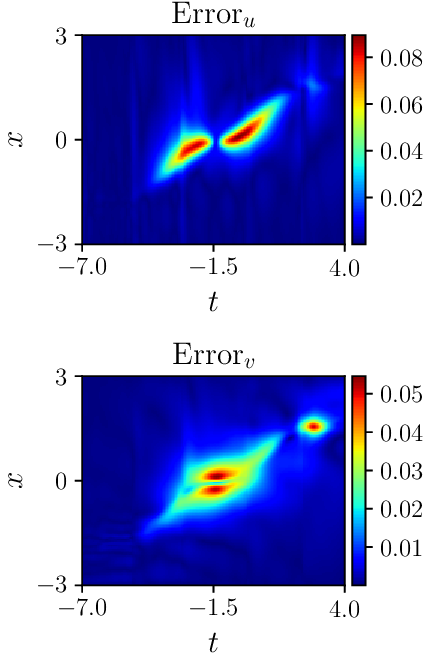}\label{figbreatheronec}}
\caption{The data-driven one-breather solution of the MT model (\ref{mt-ib}): (a) and (b) The 3D and 2D density profiles of the learned one-breather, respectively; (c) The 2D density profiles of the point-wise absolute errors between exact and learned solutions, $\mbox{Error}_u=|u-\hat{u}|$ and $\mbox{Error}_v=|v-\hat{v}|$.
\label{fig-breatherone}}
\end{figure}

As reported in Ref.\cite{chen2023tau}, the general breather of the MT model (\ref{mt01}) contains two special cases: the Akhmediev breather and the Kuznetsov-Ma breather.
For the sake of illustration, we will only consider the general one-breather and the Kuznetsov-Ma breather here.
In two cases, the parameters are taken as $\rho_1=2\rho_2=\alpha=1$, $\omega_{11}=4+2{\rm i}$, $\omega_{21}=-\frac{12}{65}+\frac{99}{65}{\rm i}$, $\zeta_{10}=0$ for the general one-breather and $\rho_1=-2\rho_2=1$, $\alpha=2$, $\omega_{11}=5+{\rm i}$, $\omega_{21}=\frac{1}{5}+{\rm i}$, $\zeta_{10}=0$ for the Kuznetsov-Ma breather, respectively.

\begin{figure}[!htbp]
\centering
\subfigure[]{\includegraphics[height=2.7in,width=1.8in]{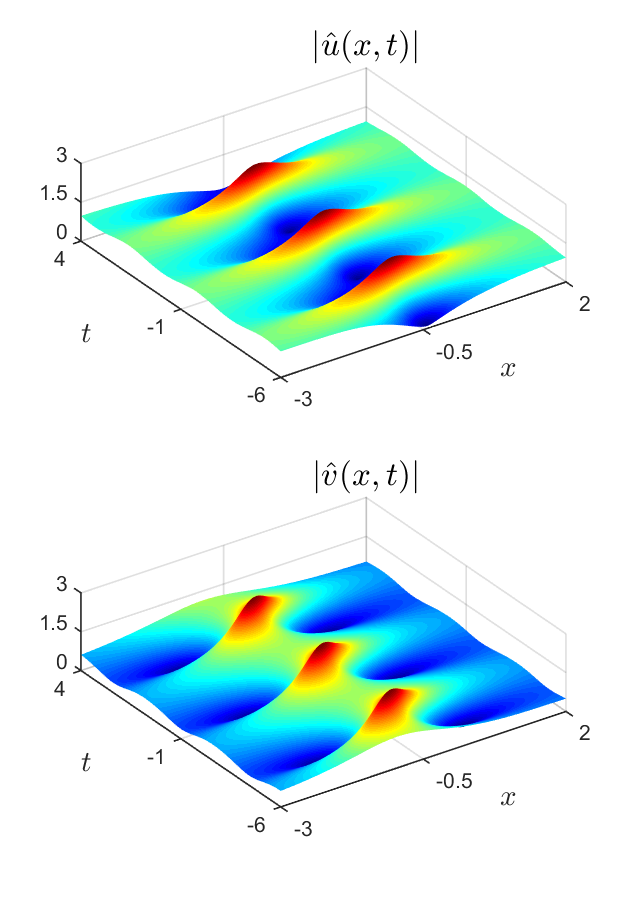}\label{figbreatherkma}}\hspace{0.2cm}
\subfigure[]{\includegraphics[height=2.7in,width=1.8in]{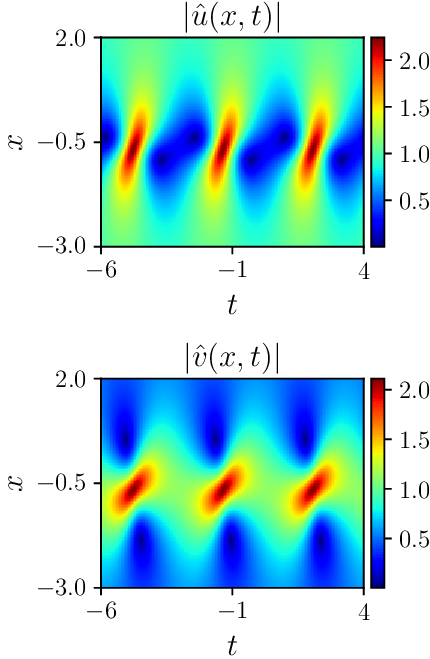}\label{figbreatherkmb}}\hspace{0.2cm}
\subfigure[]{\includegraphics[height=2.7in,width=1.8in]{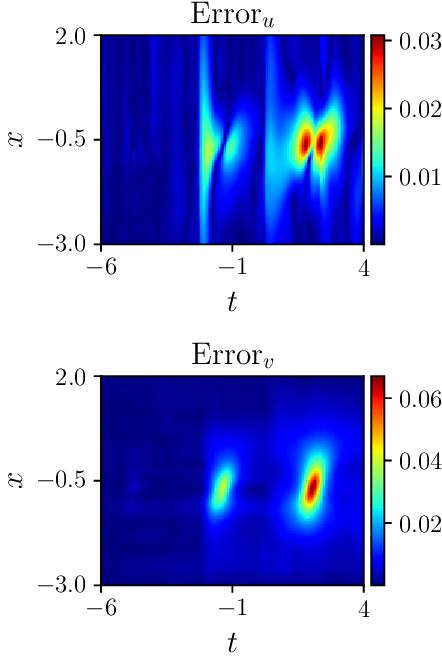}\label{figbreatherkmc}}
\caption{The data-driven Kuznetsov-Ma breather solutions of the MT model (\ref{mt-ib}): (a) and (b) The 3D and 2D density profiles of the learned Kuznetsov-Ma breather, respectively; (c) The 2D density profiles of the point-wise absolute errors between exact and learned solutions, $\mbox{Error}_u=|u-\hat{u}|$ and $\mbox{Error}_v=|v-\hat{v}|$.
\label{fig-breatherkm}}
\end{figure}

For the data-driven experiments of one-breather solutions, we set the computational domain $[L_0,L_1]\times[T_0,T_1]$ to $[-3,3]\times[-7,4]$ and $[-3,2]\times[-6,4]$ for two examples.
The initial-boundary conditions of one-breather solutions can be obtained exactly from the formulae defined in (\ref{mt-ib}).
For both types of breather solutions, we still choose $600\times1200$ grid points with the same step size to generate the data sets of the discretized initial-boundary value.
Similar to the bright and dark two-soliton solutions, we need to employ the XPINN algorithm to approximate the sophisticated breather solutions.
More specifically, the whole domains for both types of breather are first divided into 5 subdomains as shown in Tables \ref{table-breatherone} and \ref{table-breatherkm} respectively, and then the associated small interface zones are determined by defining $[\epsilon^-_k,\epsilon^+_k]=[5,3]\times \frac{T_1-T_0}{1200}$ uniformly with $k=1,2,3,4$.
The numbers of training points $N^{(k)}_{PIB}$ at each stage are given in Tables \ref{table-breatherone} and \ref{table-breatherkm}, and
the number of collocation points $N^{(k)}_{R}$ in each subdomain is set to 20000 equally, which are obtained by using the LHS method.
In each interface zone,  residual points and gradient points are taken as the corresponding part of the grid points in the discretisation of exact solution for the smooth stitching at the line $t=t_k$.
Hence the numbers of two types of sample points are given by $N^{(k)}_{IR}=N^{(k)}_{IG}=4800$.
By sequentially training the learnable parameters of each NN and combining the learned solutions throughout all subdomains,
we have both types of the predicted one-breather solutions $\hat{u}$ and $\hat{v}$ in the whole computational domain.
In Tables \ref{table-breatherone} and \ref{table-breatherkm}, we list their relative $L^2$-norm errors for $(|\hat{u}|,|\hat{v}|)$ in each subdomain and the whole domain.
Figs. \ref{fig-breatherone} and \ref{fig-breatherkm} exhibit the 3D structure and 2D density profiles of the learned solutions as well as the 2D density profiles of the point-wise absolute errors for the general one-breather and the Kuznetsov-Ma breather, respectively.
The maximum values of the absolute error for two cases are $(8.944{\rm e-}02,5.460{\rm e-}02)$ and $(3.080{\rm e-}02,6.705{\rm e-}02)$, which occur at the peaks and valleys of the local wave with the high absolute gradients.
In analogy to the bright and dark two-soliton solutions, one can see that the slight mismatches along the interface line also appear due to the different sub-networks.
As mentioned above, we can introduce more interface information to achieve the smoother stitching.
For example, these information can be obtained by adjusting the numbers of interface points, the weights of interface terms in the loss function and the sizes of interface zones adequately, or by imposing the additional smoothness conditions such as the higher-order gradients in the NN of each subdomain.

\begin{table}[htb]
\caption{Relative $L^2$-norm errors of the data-driven Kuznetsov-Ma breather solution in each subdomain and the whole domain.}
\vspace{0.2cm}
\centering
	\begin{tabular}{cccccccc}
		\hline
		\multirow{2}*{$L^2$ error} & $t$-domain & $[-6,-4.13]$ & $[-4.13,-2.25]$ & $[-2.25,0.25]$ & $[0.25, 2.33]$ & $[2.33,4]$ & $[-6,4]$  \\
		\cline{2-8}
		~ & $N_{Ib}{+}(N_{PI})$ & 1000 & 2000 & 2500 & 3000 & 3500 \\
		\hline
        $|\hat{u}|$&~&4.724e-04 & 1.713e-03 & 4.448e-03 & 5.553e-03 & 4.872e-03  & 4.011e-03\\
        $|\hat{v}|$&~&4.275e-04 & 7.680e-04 & 2.944e-03 & 5.393e-03 & 4.644e-03  & 3.458e-03\\
        \hline
	\end{tabular}
\label{table-breatherkm}
\end{table}

\subsection{Data-driven first- and second-order rogue wave solutions}

In contrast to the soliton and breather solutions shown above, the rogue wave solutions are expressed in terms of rational polynomials.
According to the derivation in \cite{chen2023rogue}, the first-order rogue wave solutions read
\begin{equation} \label{roguefirstsol}
u=\rho_1 e^{\mathrm{i}\phi} \frac{(X^*_1-\tilde{\rho} )( Y^*_1+\tilde{\rho}^* ) }{X^*_1Y^*_1+\frac{1}{4}}, \qquad
v=\rho_2 e^{\mathrm{i}\phi} \frac{(X^*_1-1)(Y^*_1+1) }{X_1Y_1+\frac{1}{4}}, \qquad
\phi=\rho \left(\frac{\rho_2}{\rho_1}x+\frac{\rho_1}{\rho_2}t\right),
\end{equation}
with
\begin{eqnarray*}
&&
X_1=\frac{\rho_2\hat{\rho}}{\rho_1}  x+\frac{\rho_1\rho\hat{\rho}}{\rho_2(\hat{\rho}+{\rm i}\rho)^2} t +\frac{1}{2}\frac{\hat{\rho}}{\hat{\rho}+{\rm i}\rho},
\qquad
Y_1=\frac{\rho_2\hat{\rho}}{\rho_1}  x+\frac{\rho_1\rho\hat{\rho}}{\rho_2(\hat{\rho}-{\rm i}\rho)^2} t -\frac{1}{2} \frac{\hat{\rho}}{\hat{\rho}-{\rm i}\rho},
\\
&&
\tilde{\rho}=\frac{\hat{\rho}}{\hat{\rho}-{\rm i}\rho_1\rho_2},\ \
\hat{\rho}=\sqrt{-\rho_1\rho_2\rho},\ \ \rho=1+ \rho_1\rho_2.
\end{eqnarray*}
Here $\rho_1$ and $\rho_2$ are arbitrary real parameters which must satisfy the condition: $-1<\rho_1\rho_2<0$.

\begin{figure}[!htbp]
\centering
\subfigure[]{\includegraphics[height=2.7in,width=1.8in]{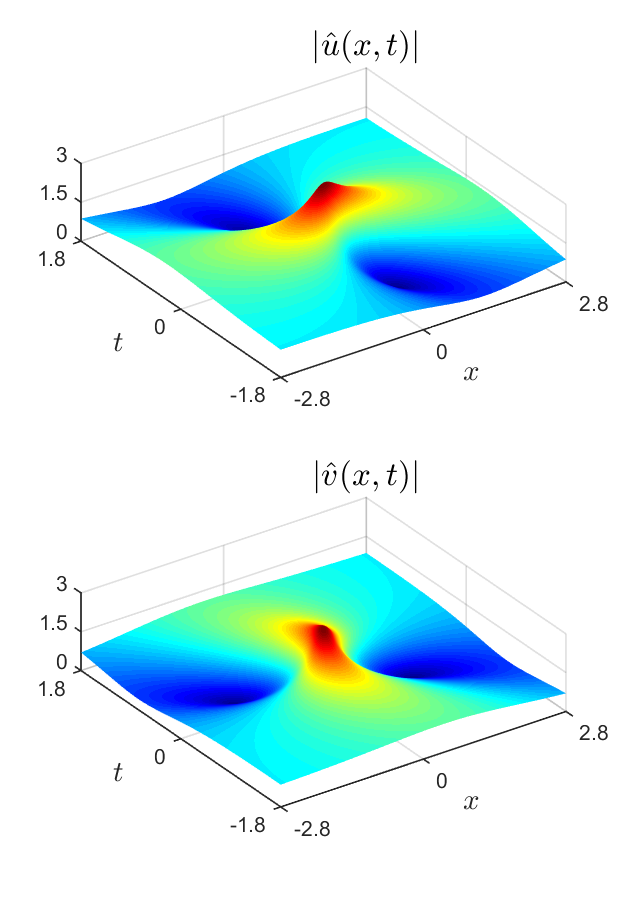}\label{figroguefirsta}}\hspace{0.2cm}
\subfigure[]{\includegraphics[height=2.7in,width=1.8in]{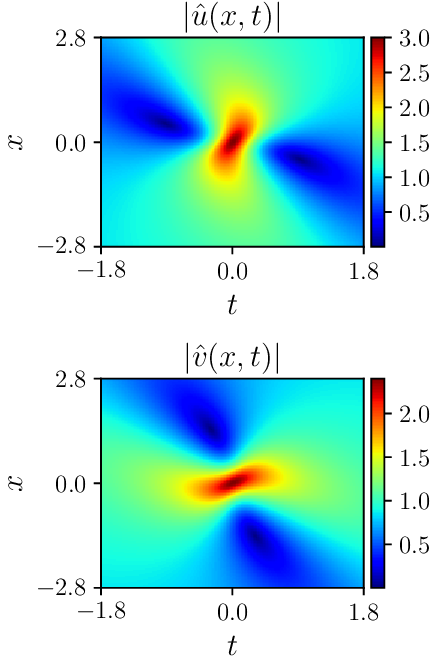}\label{figroguefirstb}}\hspace{0.2cm}
\subfigure[]{\includegraphics[height=2.7in,width=1.8in]{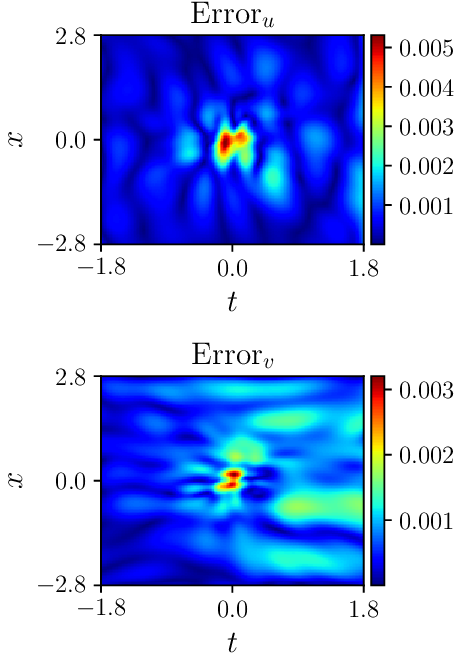}\label{figroguefirstc}}
\caption{The data-driven first-order rogue wave solution of the MT model (\ref{mt-ib}): (a) and (b) The 3D and 2D density profiles of the learned first-order rogue wave, respectively; (c) The 2D density profiles of the point-wise absolute errors between exact and learned solutions, $\mbox{Error}_u=|u-\hat{u}|$ and $\mbox{Error}_v=|v-\hat{v}|$.
\label{fig-roguefirst}}
\end{figure}

In the first-order rogue wave solutions, we take the parameters $\rho_1=1$ and $\rho_2=-\frac{4}{5}$ to perform the data-driven experiment.
For the computational domain, we take the intervals $[L_0,L_1]$ and $[T_0,T_1]$ as $[-2.8,2.8]$ and $[-1.8,1.8]$ respectively.
Then the initial and Dirichlet boundary conditions in (\ref{mt-ib}) are written explicitly, and the corresponding discretized data sets are  generated by imposing $600\times400$ grid points with the equal step length in the space-time domain.
Similar to the bright and dark one-soliton solutions, here we utilize the original PINN algorithm to simulate the simple first-order rogue wave solutions.
In the initial-boundary data sets and the whole domain, we randomly select $N_{IB}=1000$ training points and $N_R=20000$ collocation points respectively by using the LHS method.
By optimising the learnable parameters of the NN through the iterative training,
the approximated first-order rogue wave solutions $\hat{u}$ and $\hat{v}$ are successfully obtained.
In this case, the relative $L^2$-norm errors for $(|\hat{u}|,|\hat{v}|)$ are $(4.283{\rm e-}04,5.764{\rm e-}04)$.
Fig. \ref{fig-roguefirst} exhibits the 3D structure and 2D density profiles [Figs. \ref{figroguefirsta} and \ref{figroguefirstb}] of the learned first-order rogue wave solutions $(|\hat{u}|,|\hat{v}|)$, and the 2D density profiles of the point-wise absolute errors [Fig. \ref{figroguefirstc}], respectively.
It can be clearly seen that the predicted solutions agree well with the exact ones, except for the maximum absolute errors $(5.316{\rm e-}03, 3.203{\rm e-}03)$, which appear at the central peaks of rogue wave with the high gradients.

The second-order rogue wave solutions are given by \cite{chen2023rogue}
\begin{equation} \label{roguesecondsol}
u=\rho_1 e^{\mathrm{i}\phi} \frac{ g }{f^*}, \qquad
v=\rho_2 e^{\mathrm{i}\phi} \frac{ h }{ f},\qquad
\phi=\rho \left(\frac{\rho_2}{\rho_1}x+\frac{\rho_1}{\rho_2}t\right),
\end{equation}
where tau functions $f$, $g$ and $h$ are determined by
\begin{eqnarray}
f=\left\vert
\begin{array}{cc}
m^{(0,0,0)}_{11} \vspace{0.2cm}& m^{(0,0,0)}_{13} \\
m^{(0,0,0)}_{31} & m^{(0,0,0)}_{33}
\end{array}\right\vert,
\ \ \ \
g=\left\vert
\begin{array}{cc}
m^{(-1,1,0)}_{11} \vspace{0.2cm}& m^{(-1,1,0)}_{13} \\
m^{(-1,1,0)}_{31} & m^{(-1,1,0)}_{33}
\end{array}\right\vert,
\ \ \ \
h=\left\vert
\begin{array}{cc}
m^{(-1,0,1)}_{11} \vspace{0.2cm}& m^{(-1,0,1)}_{13} \\
m^{(-1,0,1)}_{31} & m^{(-1,0,1)}_{33}
\end{array}\right\vert,
\end{eqnarray}
with the elements
\begin{eqnarray*}
&&
m^{(n,k,l)}_{11}=X_1Y_1+\frac{1}{4}, \quad
m^{(n,k,l)}_{13}=X_1(Y_3+\frac{1}{6}Y^3_1)+\frac{1}{8}Y^2_1-\frac{1}{48},
\quad
m^{(n,k,l)}_{31}=Y_1(X_3+\frac{1}{6}X^3_1)+\frac{1}{8}X^2_1-\frac{1}{48}, \\
&&
m^{(n,k,l)}_{33}=(X_3+\frac{1}{6}X^3_1)(Y_3+\frac{1}{6}Y^3_1)+\frac{1}{16}(X^2_1-\frac{1}{6})(Y^2_1-\frac{1}{6})+\frac{1}{16}X_1Y_1+\frac{1}{64}.
\end{eqnarray*}
The variables $X_i$ and $Y_i$ $(i=1,3)$ are as follows:
\begin{eqnarray*}
X_{i} =\alpha_{i} x+\beta_{i} t +(n+\frac{1}{2})\theta_{i} +k \vartheta_{i} +l\zeta_r +a_{i},
\ \
Y_{i} =\alpha_{i} x+\beta^*_{i} t -(n+\frac{1}{2})\theta^*_{i} -k \vartheta^*_{i}  -l\zeta_r +a^*_{i},
\end{eqnarray*}
with the coefficients:
\begin{eqnarray*}
&& a_1=\zeta_3=0,\ \ \zeta_1=1,\ \
\alpha_1=\frac{\rho_2\hat{\rho}}{\rho_1},\ \
\alpha_3=\frac{\rho_2\hat{\rho}}{6\rho_1},\ \
\ \
\beta_1=\frac{\rho_1\rho\hat{\rho}}{\rho_2(\hat{\rho}+{\rm i}\rho)^2},\ \
\\
&&
\beta_3=\frac{\rho_1\rho(4{\rm i} \rho_1\rho_2\rho-2\hat{\rho}\rho_1\rho_2-\hat{\rho})}{6\rho_2(\hat{\rho}+{\rm i}\rho)^4},
\ \
\theta_1=\frac{\hat{\rho}}{\hat{\rho}+{\rm i}\rho},\ \
\theta_3=\frac{{\rm i}\rho^2(\rho-1+{\rm i}\hat{\rho})}{6(\hat{\rho}+{\rm i}\rho)^3},
\\
&&
\vartheta_1=\frac{\hat{\rho}}{{\rm i}(\rho-1)+\hat{\rho}},\ \
\vartheta_3=\frac{({\rm i}\rho-\hat{\rho})(\rho-1)^2}{6[{\rm i}(\rho-1)+\hat{\rho}]^3},
\ \
 \hat{\rho}=\sqrt{-\rho_1\rho_2\rho},\ \ \rho=1+ \rho_1\rho_2.
\end{eqnarray*}
Here $a_3$ is an arbitrary complex parameter, $\rho_1$,$\rho_2$ are arbitrary real parameters which need to satisfy the condition: $-1<\rho_1\rho_2<0$.

\begin{table}[htb]
\caption{Relative $L^2$-norm errors of the data-driven second-order rogue wave solution in each subdomain and the whole domain.}
\vspace{0.2cm}
	\centering
	\begin{tabular}{cccccccc}
		\hline
		\multirow{2}*{$L^2$ error} & $t$-domain & $[-3,-1.69]$ & $[-1.69,-0.38]$ & $[-0.38,1.38]$ & $[1.38, 2.83]$ & $[2.83,4]$ & $[-3,4]$  \\
		\cline{2-8}
		~ & $N_{Ib}{+}(N_{PI})$ & 1000 & 2000 & 2500 & 3000 & 3500 \\
		\hline
        $|\hat{u}|$&~&1.182e-03 & 3.587e-03 & 6.940e-03 & 4.241e-03 & 3.249e-03  & 4.739e-03\\
        $|\hat{v}|$&~&4.739e-04 & 2.813e-03 & 4.359e-03 & 4.114e-03 & 4.281e-03  & 3.573e-03\\
        \hline
	\end{tabular}
\label{table-roguesecond}
\end{table}

\begin{figure}[!htbp]
\centering
\subfigure[]{\includegraphics[height=2.7in,width=1.8in]{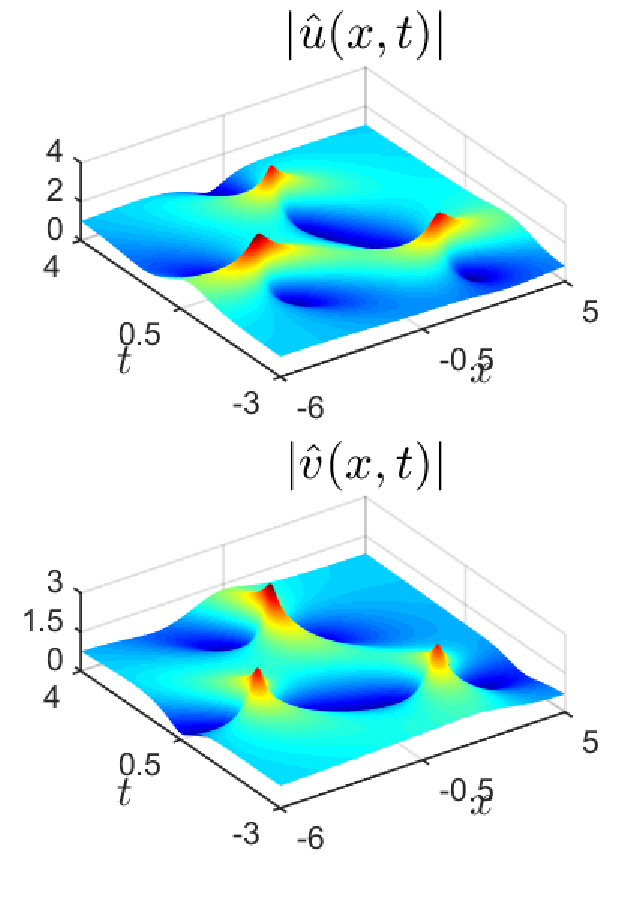}\label{figrogueseconda}}\hspace{0.2cm}
\subfigure[]{\includegraphics[height=2.7in,width=1.8in]{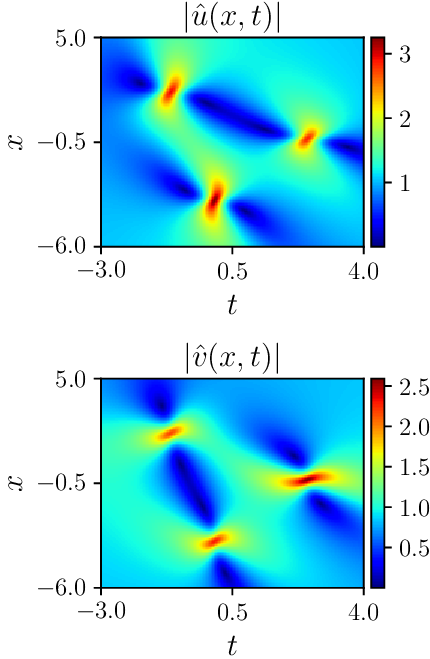}\label{figroguesecondb}}\hspace{0.2cm}
\subfigure[]{\includegraphics[height=2.7in,width=1.8in]{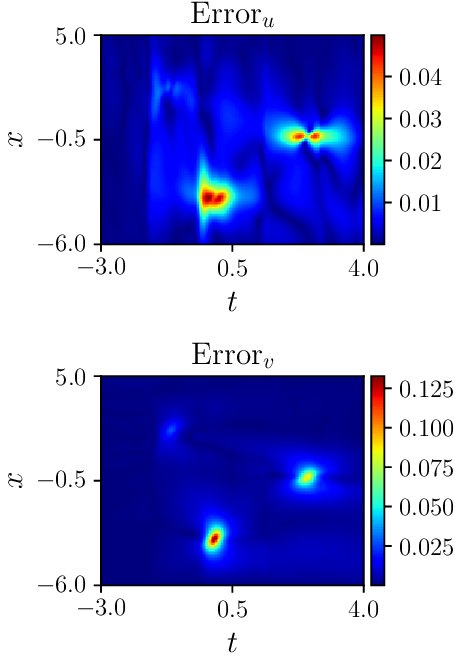}\label{figroguesecondc}}
\caption{The data-driven second-order rogue wave solution of the MT model (\ref{mt-ib}): (a) and (b) The 3D and 2D density profiles of the learned second-order rogue wave, respectively; (c) The 2D density profiles of the point-wise absolute errors between exact and learned solutions, $\mbox{Error}_u=|u-\hat{u}|$ and $\mbox{Error}_v=|v-\hat{v}|$.
\label{fig-roguesecond}}
\end{figure}

According to the Ref.\cite{chen2023rogue}, the second-order rogue wave solutions allow the wave patterns of the sole huge peak with $a_3=0$ and three peaks with $a_3\neq 0$.
Here, we only perform the data-driven simulation for the second-order rogue wave with three peaks, where we take the parameters $\rho_1=1$, $\rho_2=-\frac{4}{5}$ and $a_3=\frac{1}{2}$.

For the computational domain, the intervals $[L_0,L_1]$ and $[T_0,T_1]$ are chosen to be $[-3,2]$ and $[-6,4]$ respectively.
As defined in (\ref{mt-ib}), the initial and Dirichlet boundary conditions of the second-order rogue wave solutions can be written immediately.
It is straightforward to obtain the data sets of the discretized initial-boundary value by imposing $600\times1200$ grid points in the space-time domain with the equal step length.
In a similar way to the two-soliton and beather solutions, the complicated local features of second-order rogue wave solutions force us to use the XPINN algorithm for conducting the numerical experiments.
First, the whole domain is divided into 5 subdomains as shown in Table \ref{table-roguesecond}, and $[\epsilon^-_k,\epsilon^+_k]=[5,3]\times \frac{T_1-T_0}{1200}$ with $k=1,2,3,4$ are defined uniformly for the small interface zones.
At each stage, the numbers of training points $N^{(k)}_{PIB}$ are provided in Table \ref{table-roguesecond}, and the number of collocation points $N^{(k)}_{R}$ is equally fixed at 20000. Here we employ the LHS method to generate these random training points.
In each interface zone, residual points and gradient points are selected as the corresponding part of the grid points in the discretisation of exact solution to ensure smooth stitching at the line $t=t_k$.
Hence the numbers of two types of sample points can be obtained as $N^{(k)}_{IR}=N^{(k)}_{IG}=4800$.
By training the learnable parameters of each NN in five subdomains and stitching these predicted solutions together,
the learned second-order rogue wave solutions $\hat{u}$ and $\hat{v}$ can be constructed in the whole domain.
The relative $L^2$-norm errors for $(|\hat{u}|,|\hat{v}|)$ in each subdomain and the whole domain are presented in Table \ref{table-roguesecond}.
In Fig. \ref{fig-roguesecond}, we display the 3D structure and 2D density profiles of the learned solutions as well as the 2D density profiles of the point-wise absolute errors for the second-order rogue wave, respectively.
From Fig. \ref{figroguesecondc}, the maximum absolute errors are measured as $(4.988{\rm e-}02,1.327{\rm e-}01)$, which appear at the peaks and valleys of the rogue wave.
Particularly, unlike the two-soliton and breather cases, the relatively larger errors at the peaks and valleys make the mismatches along the interface line unnoticeable.
This is due to the fact that the small regions with higher gradients, such as the sharp peaks in Fig. \ref{figrogueseconda}, make it difficult for NN learning to capture these features.
This type of absolute error can be reduced by secondary or repetitive training by selecting more collocation points in these sharp regions.

\section{Data-driven parameters discovery of the MT model}\label{para-MT}

In this section, we consider the inverse problem of the MT model where both equations in Eq.(\ref{mt-ib}) are replaced by
\begin{equation}\label{mt-inp}
\begin{array}{l}
\mathrm{i} u_x + \lambda_1 v + \lambda_2 u|v|^2=0,\\
\mathrm{i} v_t + \lambda_3 u + \lambda_4 v|u|^2=0,
\end{array}
\end{equation}
with unknown real coefficients $\lambda_i$ for $i=1,2,3,4$.
Correspondingly, the associated complex-valued PINNs for the MT model (\ref{mt-ib}) are changed into the following form
\begin{equation}\label{mt-inpinn1}
\begin{array}{l}
f_u:=	\mathrm{i} \hat{u}_x + \lambda_1 \hat{v} + \lambda_2 \hat{u}|\hat{v}|^2,\\
f_v:=	\mathrm{i} \hat{v}_t + \lambda_3 \hat{u} + \lambda_4 \hat{v}|\hat{u}|^2.
\end{array}
\end{equation}
which are rewritten as the real-valued PINNs as follows:
\begin{equation}\label{mt-inpinn2}
\begin{array}{l}
f_p:=\hat{p}_x+\lambda_1 \hat{s}+\lambda_2 \hat{q}(\hat{r}^2+\hat{s}^2),\ \ f_q:=-\hat{q}_x+\lambda_1\hat{r} +\lambda_2\hat{p}(\hat{r}^2+\hat{s}^2), \\
f_r:=\hat{r}_t+\lambda_3\hat{q}+\lambda_4\hat{s}(\hat{p}^2+\hat{q}^2),\ \ f_s:=-\hat{s}_t+\lambda_3\hat{p}+\lambda_4\hat{r}(\hat{p}^2+\hat{q}^2),
\end{array}
\end{equation}
by taking the real-valued functions $(\hat{p},\hat{q})$ and $(\hat{r},\hat{s})$ as the real and imaginary parts of $(\hat{u},\hat{v})$, respectively.

For the inverse problem, the parameters $\lambda_i$ with $i=1,2,3,4$ in Eqs.(\ref{mt-inp}) are unknown,
some extra information of the internal region can be introduced into the PINNs of the MT model.
This implementation leads to the additional loss term defined as
\begin{equation}
MSE_{IN}=\frac{1}{N_{IN}}\sum\limits^{N_{IN}}\limits_{i=1}\left(\left|\hat{p}(x^i_{IN},t^i_{IN})-p^i_{IN}\right|^2
+ \left|\hat{q}(x^i_{IN},t^i_{IN})-q^i_{IN}\right|^2
+ \left|\hat{r}(x^i_{IN},t^i_{IN})-r^i_{IN}\right|^2
+ \left|\hat{s}(x^i_{IN},t^i_{IN})-s^i_{IN}\right|^2
\right),
\end{equation}
where $\left \{x^i_{IN},t^i_{IN}\right \}^{N_{IN}}_{i=1}$ is the set of random internal points where the number of points is $N_{IN}$,
and the discretized solutions $(p^i_{IN},q^i_{IN},r^i_{IN},s^i_{IN})$ correspond to $[p(x^i_{IN},t^i_{IN})$,$q(x^i_{IN},t^i_{IN})$,$r(x^i_{IN},t^i_{IN})$,$s(x^i_{IN},t^i_{IN})]$.
Then the new loss function in (\ref{loss1}) is modified as
\begin{equation}\label{lossin}
\mathcal{L}(\Theta)=Loss=\textbf{W}_R MSE_R + \textbf{W}_{IB} MSE_{IB} + \textbf{W}_{IN} MSE_{IN},
\end{equation}
with the weight $\textbf{W}_{IN}$.
Consequently, the redesigned PINNs of the MT model are used to learn the unknown parameters $\lambda_i$($i=1,2,3,4$) simultaneously with the solutions $(\hat{p},\hat{q},\hat{r},\hat{s})$.
It is noted that here the loss term of the initial-boundary value boundary conditions can be eliminated due to the introduction of the loss term for internal information \cite{lu2021deepxde,pang2019fpinns,lu2021constraints}.

\begin{table}[htb]
\caption{The correct MT model and the identified one with different noises driven by the bright one-soliton solutions.}
\vspace{0.2cm}
	\centering
	\begin{tabular}{cccccccc}
		\hline
		\multirow{2}*{MT model} & Item & ~ &  \\
		\cline{2-4}
		~ &  Parameters & Relative errors& \\
		\hline
        Correct & $\lambda_1=1$,\ \ $\lambda_2=1$  & $\lambda_1:0\%$,\ \ $\lambda_2:0\%$ &\\
        ~       & $\lambda_3=1$,\ \ $\lambda_4=1$  & $\lambda_3:0\%$,\ \ $\lambda_4:0\%$ & \\
        \hline
        Identified & $\lambda_1=0.9999868$,\ \ $\lambda_2=1.000026$  & $\lambda_1:0.00132\%$,\ \ $\lambda_2:0.00262\%$ &\\
         (without noise)&$\lambda_3=1.000012$,\ \ $\lambda_4=1.000082$  & $\lambda_3:0.00118\%$,\ \ $\lambda_4:0.00819\%$ &\\
        \hline
         Identified &$\lambda_1=0.9994288$,\ \ $\lambda_2=0.9972788$  & $\lambda_1:0.05712\%$,\ \ $\lambda_2:0.27212\%$ &\\
         (5 \% noise)&$\lambda_3=1.002426$,\ \ $\lambda_4=1.000761$  & $\lambda_3: 0.24258\%$,\ \ $\lambda_4: 0.07610\%$ &\\
        \hline
         Identified &$\lambda_1=0.9981983$,\ \ $\lambda_2=0.9947988$  & $\lambda_1:0.18017\%$,\ \ $\lambda_2:0.52012\%$ &\\
         (10 \% noise)&$\lambda_3=1.005139$,\ \ $\lambda_4=1.002173$  & $\lambda_3: 0.51393\%$,\ \ $\lambda_4: 0.21727\%$ &\\
        \hline
	\end{tabular}
\label{table-bightone-inverse}
\end{table}

Based on any given solutions of the MT model (\ref{mt01}), the data-driven experiment for the parameter discovery can be performed theoretically.
However, it is found from the last section for forward problems that although the complicated local solutions such as two-soliton, one-breather and second-order rogue wave can be learned with aid of the XPINN algorithm, the simple solutions  such as one-soliton and first-order rogue wave predicted by means of the PINN framework show  better accuracy with smaller absolute errors.
Hence, here we only choose three types of simple solutions including the bright, dark one-soliton and first-order rogue wave to discover unknown parameters by using the classical PINN algorithm.
For three cases, we establish the network architecture consisting of 7 hidden layers with 40 neurons per layer and fix the weights $\textbf{W}_R=\textbf{W}_{IB}=\textbf{W}_{IN}=1$ uniformly.
In these trainings, we first perform the $5000$ steps Adam optimization with the default learning rate $10^{-3}$
and then use the L-BFGS optimization with the maximum iterations $50000$.

\begin{table}[htb]
\caption{The correct MT model and the identified one with different noises driven by the dark one-soliton solutions.}
\vspace{0.2cm}
	\centering
	\begin{tabular}{cccccccc}
		\hline
		\multirow{2}*{MT model} & Item & ~ &  \\
		\cline{2-4}
		~ &  Parameters & Relative errors& \\
		\hline
        Correct & $\lambda_1=1$,\ \ $\lambda_2=1$  & $\lambda_1:0\%$,\ \ $\lambda_2:0\%$ &\\
        ~       & $\lambda_3=1$,\ \ $\lambda_4=1$  & $\lambda_3:0\%$,\ \ $\lambda_4:0\%$ & \\
        \hline
        Identified & $\lambda_1=1.000278$,\ \ $\lambda_2=0.9995433$  & $\lambda_1:0.02784\%$,\ \ $\lambda_2:0.04567\%$ &\\
         (without noise)&$\lambda_3=1.000106$,\ \ $\lambda_4=0.9999900$  & $\lambda_3:0.01057\%$,\ \ $\lambda_4:0.00100\%$ &\\
        \hline
         Identified &$\lambda_1=1.003280$,\ \ $\lambda_2=0.9984017$  & $\lambda_1: 0.32800\%$,\ \ $\lambda_2: 0.15983\%$ &\\
         (5 \% noise)&$\lambda_3=1.002683$,\ \ $\lambda_4=0.9964039$  & $\lambda_3: 0.26831\%$,\ \ $\lambda_4: 0.35961\%$ &\\
        \hline
         Identified &$\lambda_1=1.006322$,\ \ $\lambda_2=0.9959229$  & $\lambda_1: 0.63217\%$,\ \ $\lambda_2: 0.40771\%$ &\\
         (10 \% noise)&$\lambda_3=1.005789$,\ \ $\lambda_4=0.9930986$  & $\lambda_3: 0.57889\%$,\ \ $\lambda_4: 0.69014\%$ &\\
        \hline
	\end{tabular}
\label{table-darkone-inverse}
\end{table}

In the numerical experiments driven by three types of solutions, the unknown parameter $\lambda_i$ is initialized to $\lambda_i=0$, and the relative error for these unknown parameters is defined as $\frac{|\hat{\lambda}_i-\lambda_i|}{\lambda_i}$, where $\hat{\lambda}_i$ and $\lambda_i$ denote the predicted and true value respectively for $i=1,2,3,4$.
The parameters and the computational domains in the bright, dark one-soliton and first-order rogue wave solutions are still taken to be same as those for the forward problems in the last section.
By imposing $400\times600$ grid points with the equal step length in each space-time region, we obtain the discretized data sets of the initial-boundary value and the internal true solution.
Furthermore, $N_R=20000$ collocation points, $N_{IB}=1000$ initial-boundary points and $N_{IN}=2000$ internal points are generated by using the LHS method.
After optimizing the learnable parameters in the NN, the unknown parameters of the MT model in the final experimental results under three types of reference solutions are shown in Tables \ref{table-bightone-inverse}, \ref{table-darkone-inverse} and \ref{table-rogueone-inverse}.
From any of the three cases, it can be seen that the unknown parameters are correctly identified by the network model and the relative errors increase observably with the noise intensity.
In addition, whether or not the training data is affected by noise, these parameter errors are very small.
This observation is an evidence for the robustness and efficiency of the PINN approach for the MT model.

\begin{table}[htb]
\caption{The correct MT model and the identified one with different noises driven by the first-order rogue wave solutions}
\vspace{0.2cm}
	\centering
	\begin{tabular}{cccccccc}
		\hline
		\multirow{2}*{MT model} & Item & ~ &  \\
		\cline{2-4}
		~ &  Parameters & Relative errors& \\
		\hline
        Correct & $\lambda_1=1$,\ \ $\lambda_2=1$  & $\lambda_1:0\%$,\ \ $\lambda_2:0\%$ &\\
        ~       & $\lambda_3=1$,\ \ $\lambda_4=1$  & $\lambda_3:0\%$,\ \ $\lambda_4:0\%$ & \\
        \hline
        Identified & $\lambda_1=1.000048$,\ \ $\lambda_2=1.000005$  & $\lambda_1:0.00475\%$,\ \ $\lambda_2:0.00048\%$ &\\
         (without noise)&$\lambda_3=0.9999870$,\ \ $\lambda_4=0.9999554$  & $\lambda_3:0.00130\%$,\ \ $\lambda_4:0.00446\%$ &\\
        \hline
         Identified &$\lambda_1=0.9991396$,\ \ $\lambda_2=0.9976323$  & $\lambda_1: 0.08604\%$,\ \ $\lambda_2: 0.23677\%$ &\\
         (5 \% noise)&$\lambda_3=0.9989458$,\ \ $\lambda_4=0.9981177$  & $\lambda_3: 0.10542\%$,\ \ $\lambda_4: 0.18823\%$ &\\
        \hline
         Identified &$\lambda_1=0.9976596$,\ \ $\lambda_2=0.9953481$  & $\lambda_1: 0.23404\%$,\ \ $\lambda_2: 0.46519\%$ &\\
         (10 \% noise)&$\lambda_3=0.9975830$,\ \ $\lambda_4=0.9962552$  & $\lambda_3: 0.24170\%$,\ \ $\lambda_4: 0.37448\%$ &\\
        \hline
	\end{tabular}
\label{table-rogueone-inverse}
\end{table}

\section{Conclusions and discussions}

In conclusion, we investigate data-driven localized wave solutions and parameter discovery in the MT model via the PINNs and modified XPINNs algorithm.
For the forward problems, abundant localized wave solutions including soliton, breather and rogue wave are learned and compared to exact ones with relative $L^2$-norm and point-wise absolute errors.
In particular, we have modified the interface line in the domain decomposition of XPINNs into the small interface zone shared by two adjacent subdomains.
Then three types of interface condition including the pseudo initial, residual and gradient ones are imposed in the small interface zones.
This modified XPINNs algorithm is applied to learn the data-driven solutions with relatively complicated structures such as
bright-bright soliton, dark-dark soliton, dark-antidark soliton, general breather, Kuznetsov-Ma breather and second-order rogue wave.
Numerical experiments show that the improved version of XPINNs can achieve better stitching effect and faster convergence for predicting dynamical behaviors of localized waves in the MT model.
For the inverse problems, the coefficient parameters of linear and nonlinear terms in the MT model are learned with clean and noise data by using the classical PINNs algorithm.
The training results reveal that for three types of data-driven solutions, these unknown parameters can be recovered accurately even for the sample data with certain noise intensity.

\section*{Acknowledgement}

The work was supported by the National Natural Science
Foundation of China (Grant Nos. 12226332, 119251080 and 12375003)
and the Zhejiang Province Natural Science Foundation of China (Grant No. 2022SJGYZC01).

\end{document}